\input epsf
\documentstyle{amsppt}
\pagewidth{5.4truein}\hcorrection{0.55in}
\pageheight{7.5truein}\vcorrection{0.75in}
\TagsOnRight
\NoRunningHeads
\catcode`\@=11
\def\logo@{}
\footline={\ifnum\pageno>1 \hfil\folio\hfil\else\hfil\fi}
\topmatter
\title A triangular gap of size two \\ in a sea of dimers on a $60^\circ$ angle  
\endtitle
\endtopmatter
\document

\def\mysec#1{\bigskip\centerline{\bf #1}\message{ * }\nopagebreak\bigskip\par}

\def\myref#1{\item"{[{\bf #1}]}"}

\def\pf{{\it Proof.\ }}

\def\epf{\hfill{$\square$}\smallpagebreak}

\def\cite#1{\relaxnext@
  \def\nextiii@##1,##2\end@{[{\bf##1},\,##2]}%
  \in@,{#1}\ifin@\def\next{\nextiii@#1\end@}\else
  \def\next{[{\bf#1}]}\fi\next}
\def\proclaimheadfont@{\smc}

\def\pf{{\it Proof.\ }}

\define\Z{{\bold Z}}

\define\M{\operatorname{M}}
\define\de{\operatorname{d}}
\define\q{\operatorname{q}}

\define\twoline#1#2{\line{\hfill{\smc #1}\hfill{\smc #2}\hfill}}
\define\twolinetwo#1#2{\line{{\smc #1}\hfill{\smc #2}}}
\define\twolinethree#1#2{\line{\phantom{poco}{\smc #1}\hfill{\smc #2}\phantom{poco}}}

\def\mypic#1{\epsffile{#1}}

\bigskip
\bigskip
\twolinethree{{\smc Mihai Ciucu\footnote{Supported in part by NSF grant DMS-1101670.}}}{{\smc Ilse Fischer\footnote{Supported in part by Austrian Science Foundation FWF, START grant Y463.}}}

\bigskip
\twolinethree{{\rm Indiana University}}{{\rm \rm Universit\"at Wien}}
\twolinethree{{\rm Department of Mathematics}}{{\rm Fakult\"at f\"ur Mathematik}}
\twolinethree{{\rm Bloomington, IN 47401, USA}}{{\rm Nordbergstra\ss e 15}}
\twolinethree{{\rm }}{{\rm A-1090 Wien, Austria}}



\bigskip
\bigskip
{\eightpoint {\smc Abstract.} {\rm We consider a triangular gap of side two in a $60^\circ$ angle on the triangular lattice whose sides are zig-zag lines. We study the interaction of the gap with the corner as the rest of the angle is completely filled with lozenges. We show that the resulting correlation is governed by the product of the distances between the gap and its five images in the sides of the angle. This provides a new aspect of the parallel between the correlation of gaps in dimer packings and electrostatics developed by the first author in previous work. }}

\bigskip
\bigskip

\define\And{1}
\define\fiveandahalf{2}
\define\pptwo{2}
\define\ri{3}
\define\sc{4}
\define\ec{5}
\define\ef{6}
\define\ov{7}
\define\free{8}
\define\gd{9}
\define\DT{10}
\define\FS{11}
\define\GV{12}
\define\Kuo{13}
\define\Kup{14}
\define\Lind{15}
\define\MRR{16}
\define\PBM{17}
\define\Sta{18}
\define\Ste{19}


\define\eba{2.1}
\define\ebb{2.2}
\define\ebc{2.3}
\define\ebd{2.4}
\define\ebe{2.5}

\define\eca{3.1}
\define\ecb{3.2}
\define\ecc{3.3}
\define\ecd{3.4}
\define\ece{3.5}
\define\ecf{3.6}
\define\ecg{3.7}
\define\ech{3.8}
\define\eci{3.9}
\define\ecj{3.10}

\define\eda{4.1}
\define\edb{4.2}
\define\edc{4.3}

\define\eea{5.1}
\define\eeb{5.2}
\define\eec{5.3}
\define\eed{5.4}
\define\eee{5.5}
\define\eef{5.6}
\define\eeg{5.7}
\define\eeh{5.8}

\define\efa{6.1}
\define\efb{6.2}
\define\efc{6.3}
\define\efd{6.4}
\define\efe{6.5}
\define\eff{6.6}
\define\efg{6.7}

\define\ega{7.1}
\define\egb{7.2}
\define\egc{7.3}
\define\egd{7.4}
\define\ege{7.5}
\define\egf{7.6}
\define\egg{7.7}
\define\egh{7.8}
\define\egi{7.9}
\define\egj{7.10}
\define\egk{7.11}
\define\egl{7.12}
\define\egm{7.13}
\define\egn{7.14}

\define\eha{8.1}
\define\ehb{8.2}
\define\ehc{8.3}
\define\ehd{8.4}
\define\ehe{8.5}


\define\tba{2.1}
\define\tbb{2.2}

\define\tca{3.1}
\define\tcb{3.2}

\define\tda{4.1}

\define\tea{5.1}

\define\tfa{6.1}

\define\tga{7.1}


\define\fbc{2.3}
\define\fbd{2.4}
\define\fbe{2.5}

\define\fca{3.1}
\define\fcb{3.2}
\define\fcc{3.3}
\define\fcd{3.4}
\define\fce{3.5}

\define\fea{5.1}

\vskip-0.05in
\mysec{1. Introduction}

In their paper \cite{\FS} from 1963, Fisher and Stephenson have introduced the concept of the correlation of two monomers in a sea of dimers, and based on their precise numerical findings conjectured that this correlation is rotationally invariant in the scaling limit. In a series of articles (see \cite{\ri}\cite{\sc}\cite{\ec}\cite{\gd}), the first author has extended the problem of Fisher and Stephenson to the situation when one is allowed to have any finite number of gaps, each of an arbitrary size, and has shown that a close parallel to electrostatics emerges: As the distances between the gaps approach infinity, their correlation is given by the exponential of the electrostatic energy of a two dimensional system of charges that correspond to the gaps in a natural way.

This parallel to electrostatics has been extended in \cite{\ef} and \cite{\ov}, where it was shown that the discrete field of the average tile orientations approaches, in the scaling limit, the electric field. 

One particular aspect of this analogy is the behavior of the correlation of gaps near the boundary of lattice regions, which turns out to be in close connection with the behavior of charges near conductors. In \cite{\sc} it was shown that the asymptotics of the correlation of gaps on the triangular lattice near a constrained zig-zag boundary is given by a variant of the method of images from electrostatics, in which the image charges have the same signs as the original ones. The case of a free boundary was considered in \cite{\free}, where it was shown that the correlation of a single gap of size two with a free lattice line boundary on the triangular lattice is given, in the scaling limit, precisely by the method of images from electrostatics.

In this paper the analogy to the method of images is given more substance by establishing it in a more complex setting, in which the gap has not just one image (as it was the case in \cite{\sc} and \cite{\free}), but five. Indeed, we consider a triangular gap of size two in a $60^\circ$ degree angular region on the triangular lattice whose sides are zig-zags. The gap has two direct images in the two sides of the angular region, which generate further images in the sides, to a total of five images of the original gap. The main result of this paper (see Theorem {\tba}) is that the asymptotics of the correlation of the gap with the corner of the angular region, as the distances between the gap and the sides grow large, is given by a numerical constant times the exponential of one sixth of the electrostatic energy of the 2D system of charges consisting of the gap viewed as a charge, together with its above five images.

We note that, from the point of view of the literature on plane partitions and their symmetry classes (see for instance \cite{\Sta}, \cite{\And}, \cite{\Ste} and \cite{\Kup}), parts (b) and (c) of Proposition {\tca} of this paper represent generalizations of the cyclically symmetric, self-complementary case, first solved by Kuperberg in \cite{\Kup} (see \cite{\fiveandahalf} for a simple proof).

\mysec{2. Statement of the main result and physical interpretation}

Let $n\geq2$ and $x\geq0$ be integers. Consider the pentagonal region illustrated by Figure~{\tba}, where the top side has length $x$, the southeastern side has length $n-4$, and the western and northeastern sides follow zig-zag lattice paths of lengths $2n-4$ and $2n$, respectively. Denote the resulting region by $D_{n,x}$.

For  positive integers $R$ and $v$, we define $D_{n,x}(R,v)$ to be the region obtained from $D_{n,x}$ by removing from it the up-pointing lattice triangle of side-length 2 positioned as indicated in Figure {\tbb}.

Let $D_{n,x}^0$ be the region obtained from $D_{n,x}$ by removing the up-pointing unit triangles that fit in its first and third ``bumps'' from the top along the northeastern side (Figure {\fbd} illustrates the case $n=11$, $x=1$).

For $x=1$ and fixed $R$ and $v$, as $n$ grows to infinity the gap is effectively in an infinite angular region whose sides are zig-zag lattice paths meeting at a $60^\circ$ angle. Define the correlation $\omega(R,v)$ of the gap with the corner of this angle by
$$
\omega_c(R,v):=\lim_{n\to\infty}\frac{\M(D_{n,1}(R,v))}{\M(D_{n,1}^0)},
\tag\eba
$$
where, for a lattice region $D$ on the triangular lattice, $\M(D)$ denotes the number of lozenge tilings of $D$ (the subscript $c$ indicates that the correlation feels the interaction with the corner of the angle, as $R$ and $v$ are fixed). 
The particular denominator above was chosen because it turns out to be a convenient normalizing factor (see for instance Lemma {\tea}); note that the seemingly simpler choice of removing the up-pointing unit triangles from the top two bumps does not work, as the resulting region does not have any lozenge tilings.
In the special case $R=4$, $v=5$ and $n=11$, the regions at the numerator and denominator on the right hand side of (\eba) are shown in Figures {\fbc} and~{\fbd}, respectively.

\topinsert
\twoline{\mypic{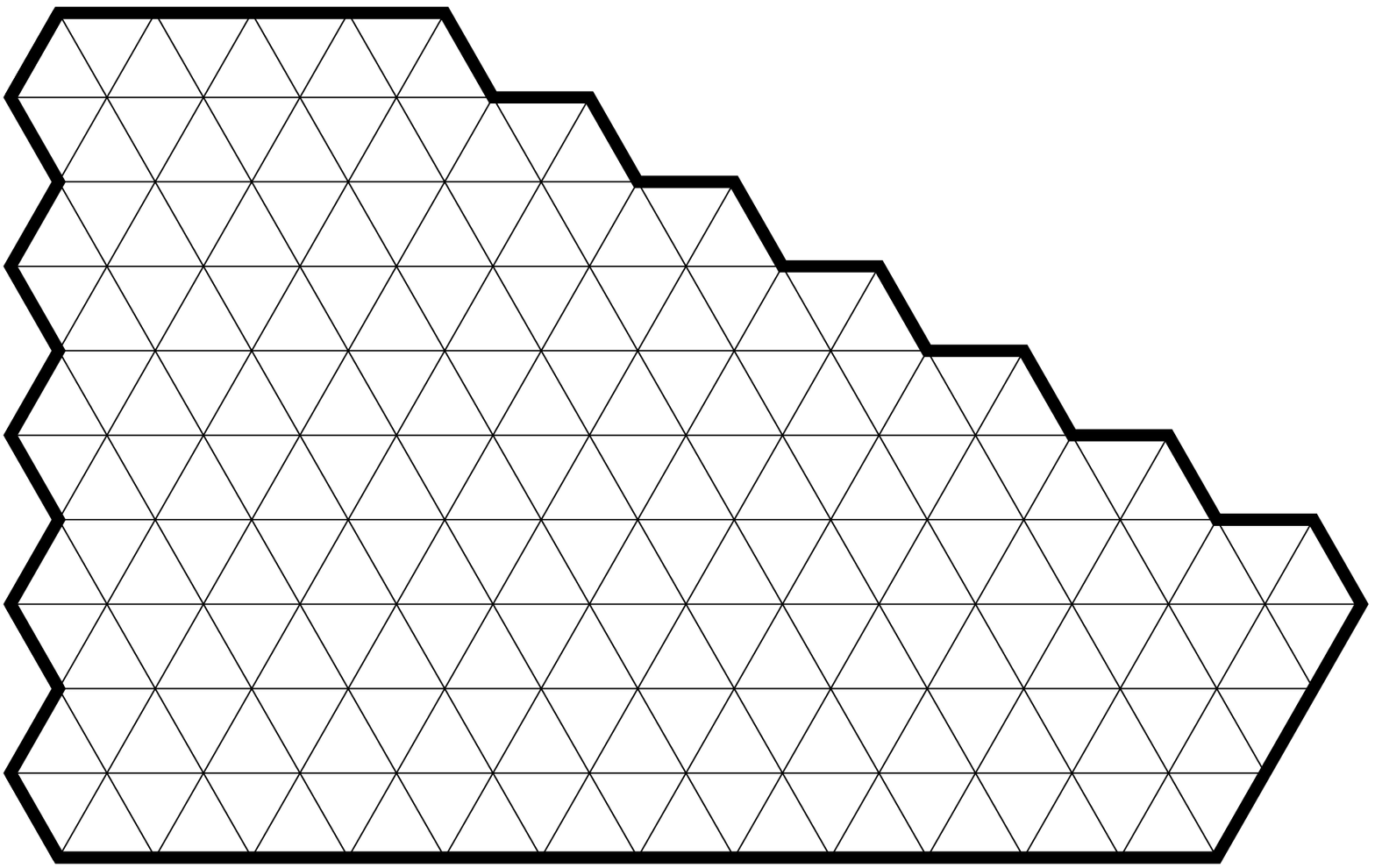}}{\mypic{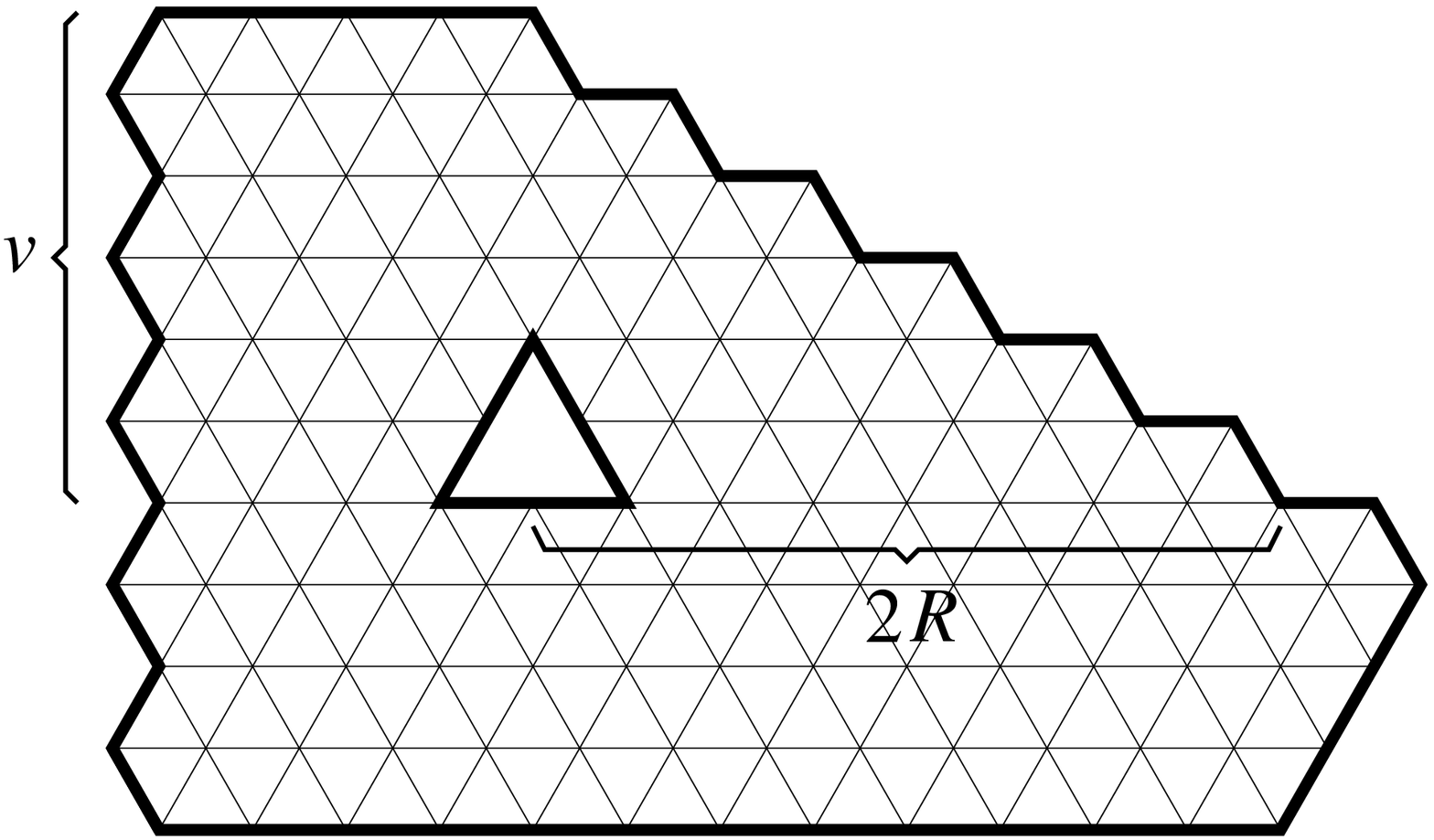}}
\twoline{Figure~{\tba}. {\rm $D_{7,4}$.}}
{Figure~{\tbb}. {\rm  $D_{7,2}(4,3)$.}}
\endinsert

\topinsert
\twoline{\!\!\!\!\mypic{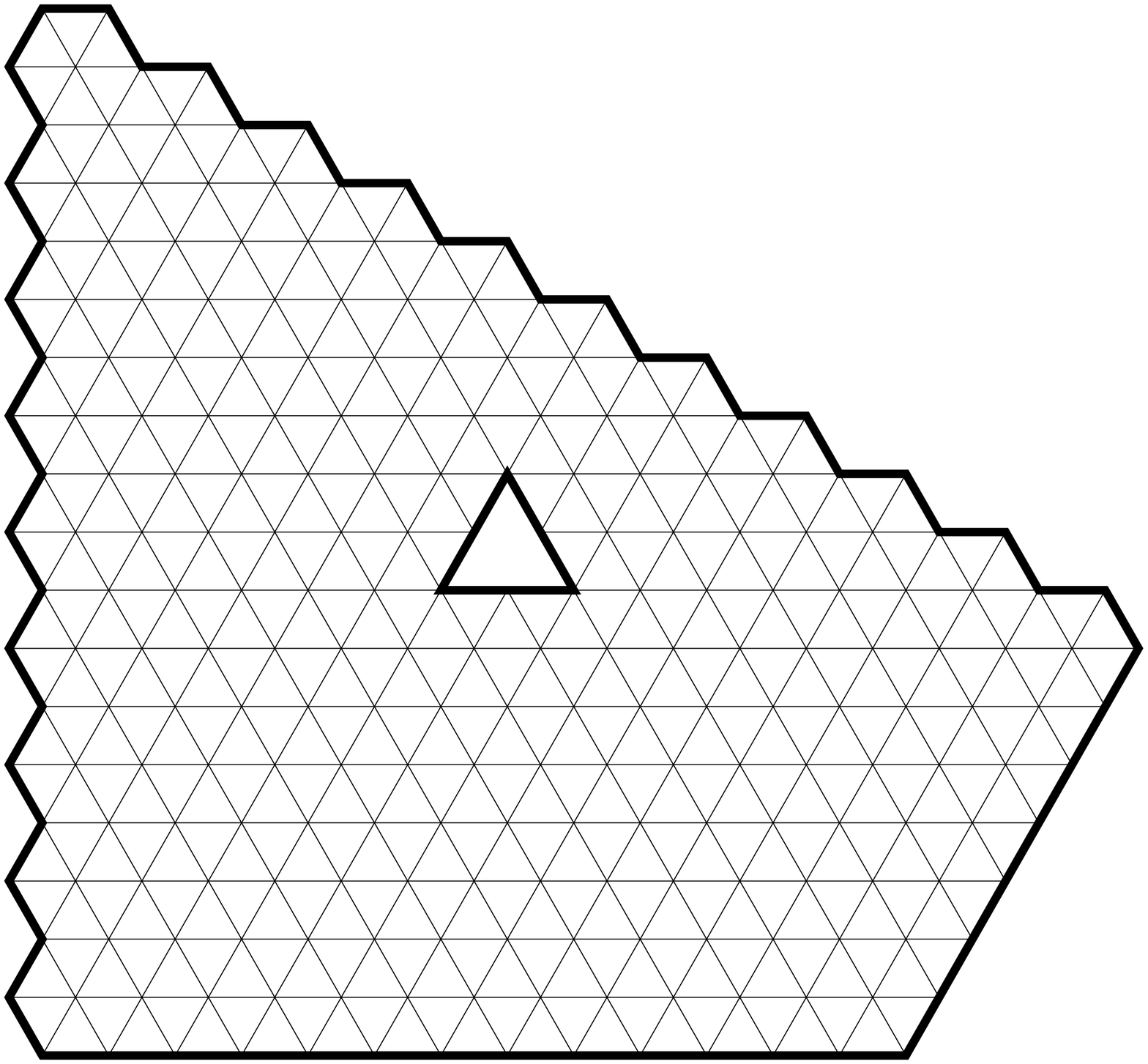}}{\ \ \ \ \ \ \mypic{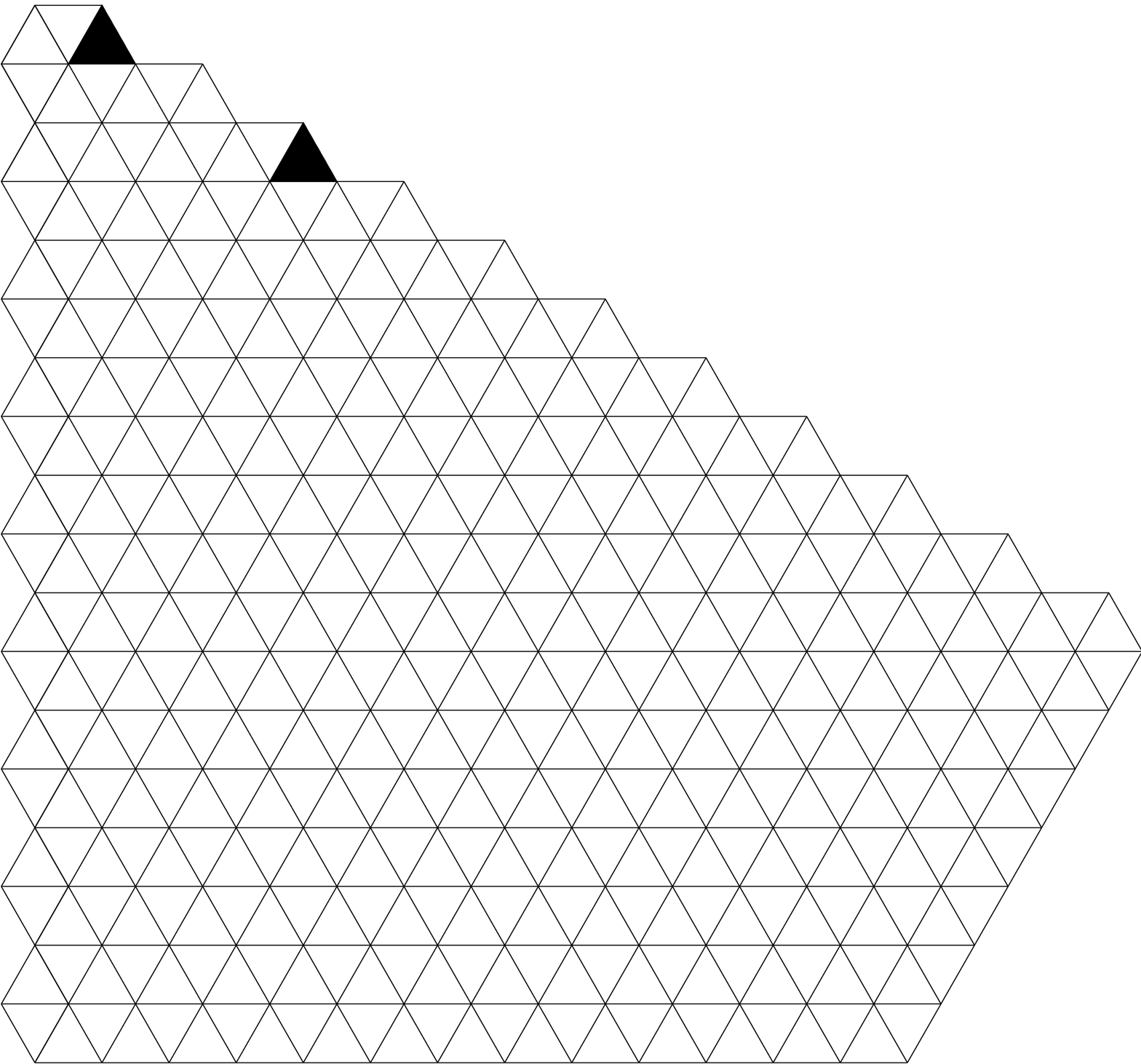}}
\twoline{Figure~{\fbc}. {\rm $D_{11,1}(4,5)$.}}
{Figure~{\fbd}. {\rm $D_{11,1}^0$.}}
\endinsert



\topinsert
\centerline{\mypic{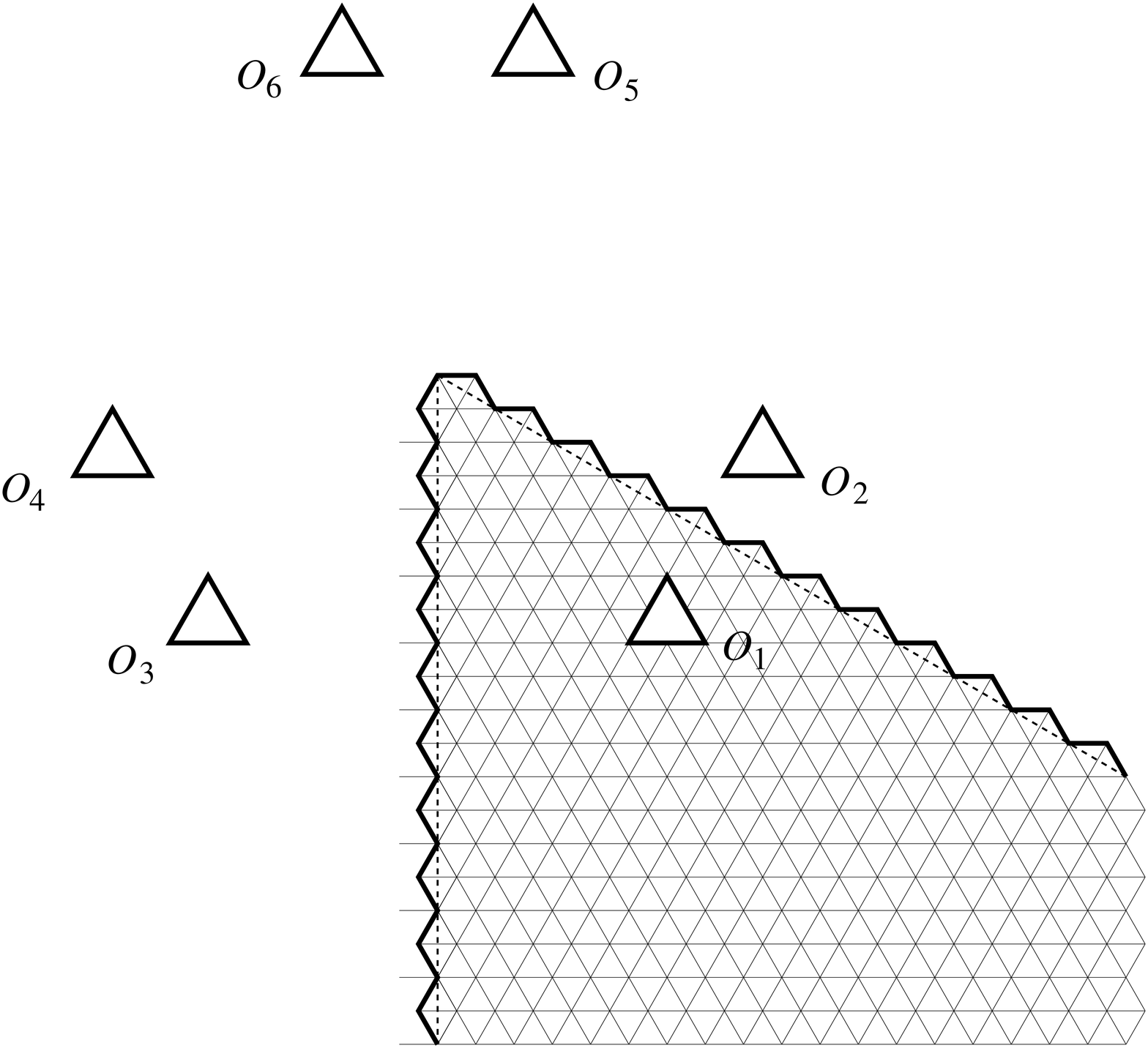}}
\centerline{{\smc Figure~{\fbe}. {\rm  The gap and its five images for $R=3$, $v=4$.}}}
\endinsert

The main result of this paper is the following.

Let us denote our triangular gap of side two inside the $60^\circ$ angle by $O_1$, and let $\ell_1$ and $\ell_2$ be the straight lines supporting the zig-zag lattice paths that form the northeastern and western sides of our $60^\circ$ angular region, respectively (they are indicated by dashed lines in Figure {\fbe}). Let $O_2$ and $O_3$ be the mirror images of $O_1$ in $\ell_1$ and $\ell_2$, respectively. Let $O_4$ be the mirror image of $O_2$ in $\ell_2$, and $O_5$ the mirror image of $O_3$ in $\ell_1$. Then the mirror image of $O_4$ in $\ell_1$ is the same as the mirror image of $O_5$ in $\ell_2$; denote it by $O_6$. (Note that $\{O_1,O_2,O_3,O_4,O_5\}$ is the set of all images $O_1$ would see if the sides of the angle were mirrors.)

\proclaim{Theorem \tba} As $R$ and $v$ approach infinity, we have
$$
\spreadlines{4\jot}
\align
\omega_c(R,v)
&
\sim
\frac{4}{81}
R(3v-R)(3v-2R)(R^2-3Rv+3v^2)
\\
&
\sim
\frac{1}{1944}\,
\root{3}\of{\prod_{1\leq i<j\leq 6} \de(O_i,O_j)},
\tag\ebb
\endalign
$$
where $\de$ is the Euclidean distance.

\endproclaim 

\flushpar
{\smc Remark 1.} In \cite{\ec} the first author has shown that if $O_1,\dotsc,O_n$ are unions of collinear triangles of side two (which can point up or down, but are of the same kind within each $O_i$), then, for large separations between the $O_i$'s, the asymptotics of their correlation in the bulk is given by
$$
\omega(O_1,\dotsc,O_n)\sim c \prod_{1\leq i<j \leq n} \de(O_i,O_j)^{\frac12 \q(O_i)\q(O_j)},
\tag\ebc
$$
where $\q(O)$ denotes the charge of the gap $O$, defined to be the number of up-pointing unit triangles in $O$ minus the number of down-pointing unit triangles in $O$, and the multiplicative constant $c$ depends only on the structure of the individual gaps, and not on their relative position.

Note that, since our gap and its five images all have charge equal to 2, the asymptotics (\ebb) of the correlation of the gap in our $60^\circ$ angle can be written as
$$
\frac{1}{1944}\,
\root{6}\of{\prod_{1\leq i<j\leq 6} \de(O_i,O_j)^{\frac12 \q(O_i)\q(O_j)}}.
\tag\ebd
$$
Thus, using (\ebc), one can rewrite the statement of Theorem {\tba} as
$$
\omega_c(O_1)\sim c'\root{6}\of{\omega(O_1,\dotsc,O_6)},
\tag\ebe
$$
where $c'$ is some explicit numerical constant. Thus the correlation at the corner can be expressed in terms of the correlation in the bulk of the gap with its images, much like in the method of images of electrostatics, when the electric field created by a charge near a conductor can be found by replacing the conductor with a suitable system of image charges. This is the main point we make in this paper in terms of physical interpretation, taking the electrostatic analogy developed in \cite{\sc}\cite{\ec}\cite{\ef}\cite{\ov}\cite{\free} one step further.

The case of triangular gaps of size two near a zig-zag boundary was seen in \cite{\sc} to be given by a formula analogous to (\ebe), in which one takes the correlation of the collection of gaps together with their mirror images, and extracts the square root from it. The case of a single triangular gap of size two near an open lattice line boundary was treated in \cite{\free}, where it was shown that the correlation has asymptotics given by the square root of the pair of gaps consisting of the original gap and its mirror image in the boundary. Formula (\ebe) adds to this small collection of known cases a new, more complex instance, and clearly suggests a conjectural answer in a variety of similar circumstances.

%

\mysec{3. Exact formulas for regions with one or two dents}

\topinsert
\centerline{\mypic{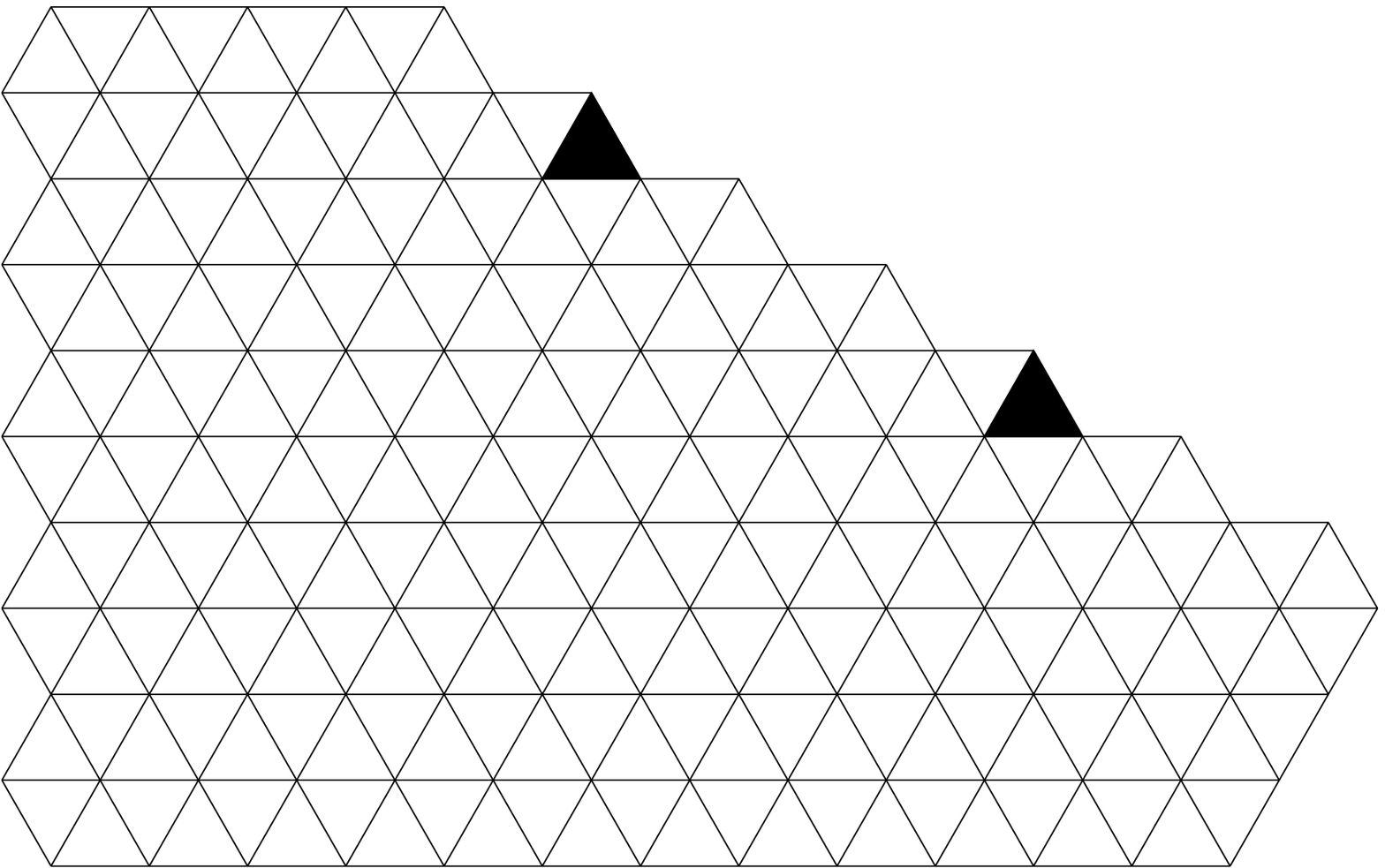}}
\centerline{{\smc Figure~{\fca}. {\rm  $E_{7,4}(2,5)$.}}}
\endinsert

The region $D_{n,x}$ has $n$ ``bumps'' along its northeastern side. Number them from 1 to $n$ starting from top. For any integers $1\leq i<j\leq n$, define $E_{n,x}(i,j)$ to be the region obtained from $D_{n,x}$ by removing the up-pointing unit triangles at the $i$th and $j$th bumps; in particular, $E_{n,1}(1,3)$ is the same as the region $D_{n,1}^0$ used for the normalizing factor in the definition (\eba) of the correlation of our triangular gap (Figure {\fca} illustrates the case $n=7$, $x=4$, $i=2$ and $j=5$).

In order to determine the number of lozenge tilings of the regions $E_{n,x}(i,j)$ (which will be needed in our proof of Theorem {\tba}), we need to consider the following two additional families of regions. 

For integers $n\geq1$, $x\geq0$ and $1\leq i \leq n$, let $F_{n,x}(i)$ be the region described in Figure~{\fcb}, where the top side has length $x$, the southeastern side has length $n-2$, the western and northeastern sides follow zig-zag paths of lengths $2n-2$ and $2n$, respectively, and the up-pointing unit triangle at the $i$th bump from the top on the northeastern side has been removed (note that $F_{1,x}(1)$ is the empty set).

Finally, for non-negative integers $n$ and $x$, let $G_{n,x}$ be the region defined by Figure {\fcc}, where the top side has length $x$, the southeastern side has length $n$, and the western and northeastern sides follow zig-zag paths of lengths $2n$.

The number of lozenge tilings of these regions is given by the following result.
Recall that for non-negative integers $k$, the Pochhammer symbol $(a)_k$ is defined by
$$
(a)_k:=a(a+1)\cdots(a+k-1).\tag\eca
$$
We will also use the standard extension of the Pochhammer symbol to a negative integer index, defined by
$$
(a)_{-k}:=\frac{1}{(a-k)(a-k+1)\cdots(a-1)},\tag\ecb
$$
where $k$ is a positive integer.

\proclaim{Proposition \tca} $(${\text{\rm a}}$)$. For any non-negative integers 
$n$ and $x$ we have
$$
\M(G_{n,x})=\frac{1}{2^n}\prod_{k=1}^n
\frac{(2x+2k)_k\,\left(x+2k+\frac12\right)_{k-1}}
         {(k)_k\,\left(x+k+\frac12\right)_{k-1}}.
\tag\ecc
$$

 $(${\text{\rm b}}$)$. For any integers $n\geq1$ and $x\geq0$, we have for any $i=1,\dotsc,n$ that
$$
\M(F_{n,x}(i))=
\M(G_{n-2,x+3})
\frac{(x+1)_{i-1}\,(n-i+1)_{i-1}\,(2x+2n+i)_{i-2}(2x+2n)}
     {(2n-i)_{i-1}\,(n+x)_{i-1}\,(i-1)!},
\tag\ecd
$$
where for $m<0$ the region $G_{m,x}$ is defined to be the empty set. 

 $(${\text{\rm c}}$)$. For any integers $n\geq2$ and $x\geq0$, we have for any $1\leq i<j\leq n$ that
$$
\M(E_{n,x}(i,j))=
\frac{\M(F_{n-1,x}(i))\M(F_{n,x}(j))-\M(F_{n-1,x}(j))\M(F_{n,x}(i))}
     {\M(G_{n-1,x})}.
\tag\ece
$$

\endproclaim

\topinsert
\twoline{\mypic{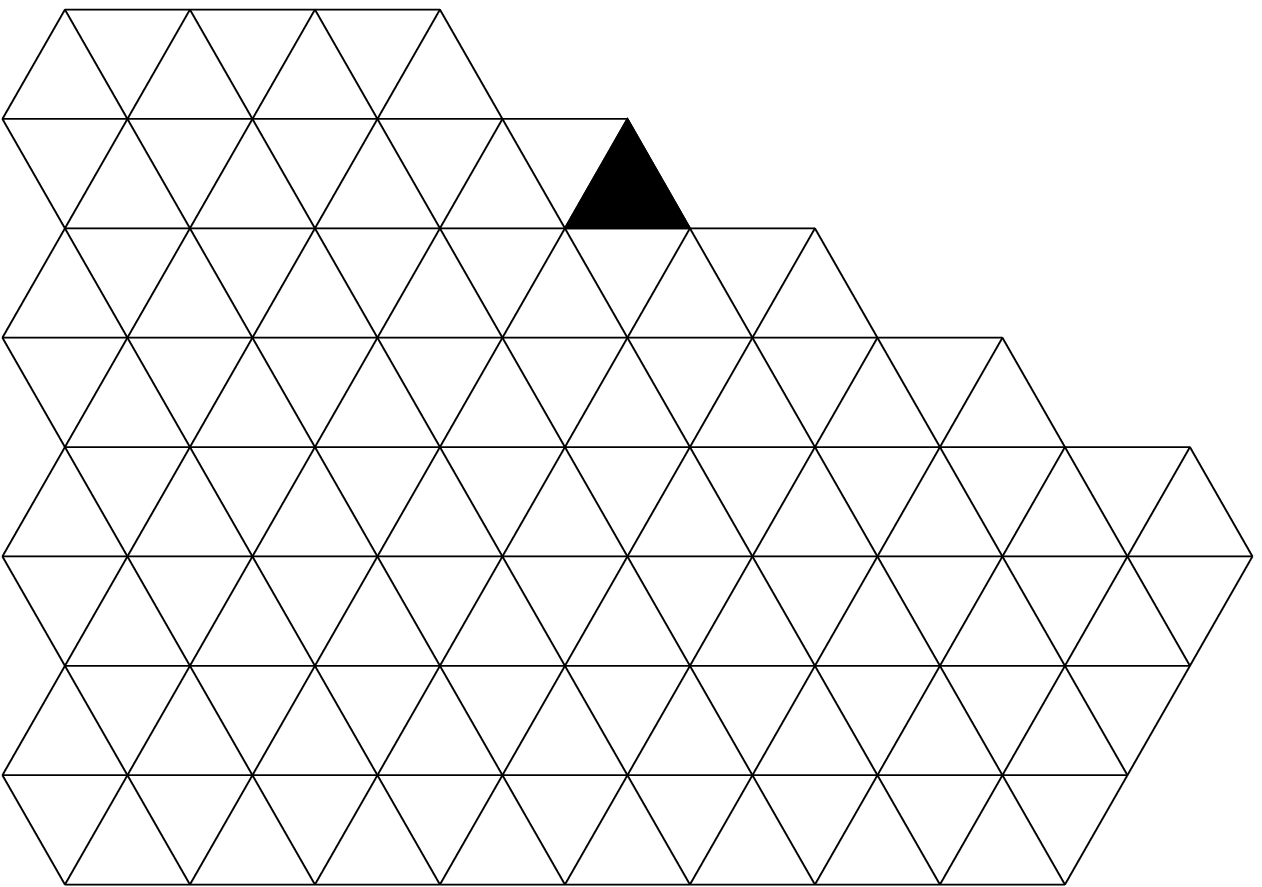}}{\mypic{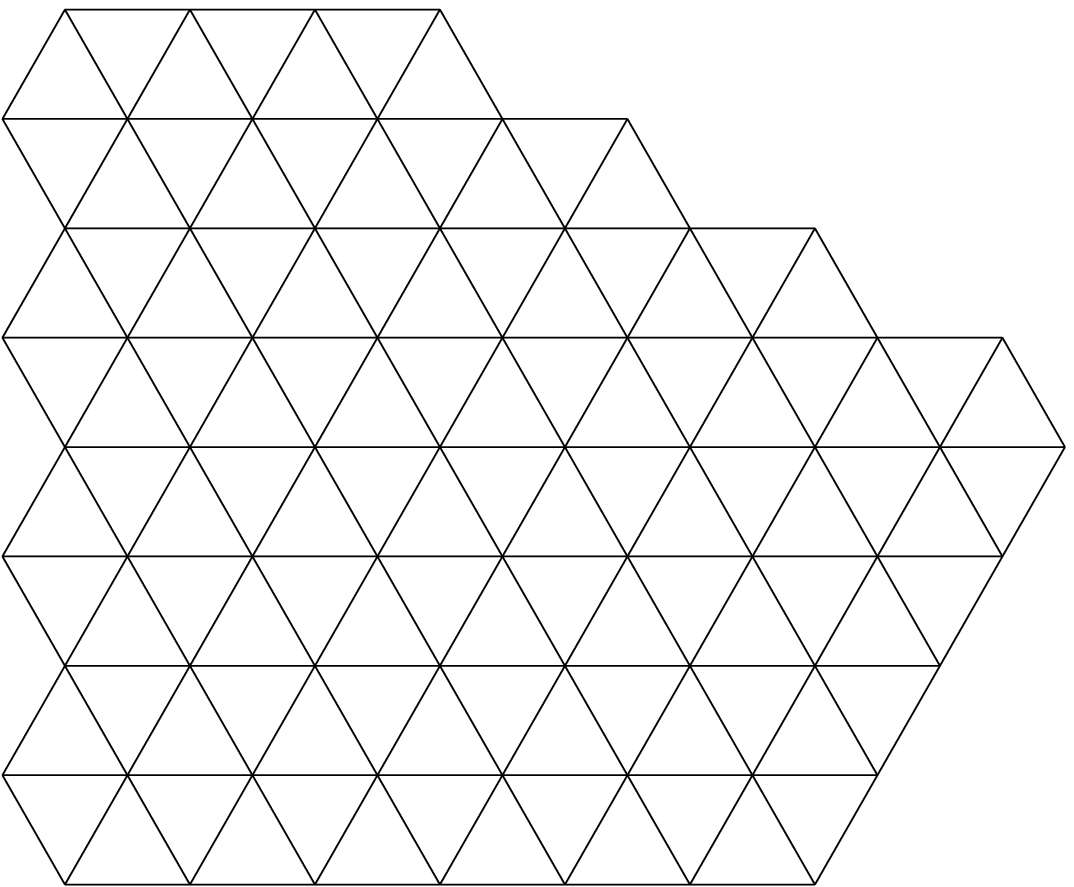}}
\twoline{Figure~{\fcb}. {\rm $F_{5,3}(2)$.}}
{Figure~{\fcc}. {\rm  $G_{4,3}$.}}
\endinsert

In the proof of the above result we make essential use of Kuo's powerful graphical condensation method (see \cite{\Kuo}). For ease of reference, we state below the particular instance of Kuo's general results that we need for our proofs (which is Theorem~2.4 in \cite{\Kuo}).

\proclaim{Theorem {\tcb} (Kuo)} Let $G=(V_1,V_2,E)$ be a plane bipartite graph in which $|V_1|=|V_2|+1$. Let vertices $a$, $b$, $c$ and $d$ appear cyclically on a face of $G$. If $a,b,c\in V_1$ and $d\in V_2$, then
$$
\M(G-b)\M(G-\{a,c,d\})=\M(G-a)\M(G-\{b,c,d\})+\M(G-c)\M(G-\{a,b,d\}).
\tag\ecf
$$

\endproclaim

{\it Proof of Proposition 3.1.} (a). This part follows directly from results in \cite{\pptwo} and \cite{\MRR}. Indeed, it is readily seen that if the forced lozenges are removed from the region $C_{n,x}$ defined in \cite{\pptwo,\S4}, then the remaining region is precisely $G_{n-1,x+1}$. Therefore
$$
\M(G_{n,x})=\M(C_{n+1,x-1}).\tag\ecg
$$
Furthermore, by \cite{\pptwo,(4.7)}, we have that
$$
\M(C_{n,x})=\det\left({x+i+j\choose 2j-i}\right)_{0\leq i,j\leq n-1}.\tag\ech
$$
In turn, by \cite{\MRR, Theorem\,7}, the determinant above evaluates as
$$
\det\left({x+i+j\choose 2j-i}\right)_{0\leq i,j\leq n-1}=
\frac{1}{2^{n-1}}\prod_{k=1}^{n-1}
\frac{(2x+2k+2)_k\,\left(x+2k+\frac32\right)_{k-1}}
         {(k)_k\,\left(x+k+\frac32\right)_{k-1}}.\tag\eci
$$
Then (\ecc) follows by (\ecg)--(\eci).

(b). We prove this part by induction on $n$. For $n=1$, the only choice for $i$ is $i=1$. Since by definition $F_{1,x}(1)=\emptyset$, we have $\M(F_{1,x}(1))=1$, which agrees with the right hand side of (\ecd), due to (\ecb) and the convention in the statement, by which $G_{-1,x+3}=\emptyset$.

For $n=2$, there are two choices for $i$, $i=1$ and $i=2$. For $i=1$, the region on the left in (\ecd) is $F_{2,x}(1)$, in which all lozenges are forced; so the left hand side of (\ecd) is~1. On the other hand, the $G$ region on the right hand side of (\ecd) is empty in this case (thus contributing a factor of 1), while the ratio on the right hand side of (\ecd) also equals~1, due to the definition (\ecb). It is easy to check that for $i=2$ both sides of (\ecd) equal $x+1$: for the right hand side this is immediate, and for the left hand side it follows as $F_{2,x}(2)$ has one forced lozenge, after the removal of which one is left with a hexagon having side-lengths 1, 1, $x$, 1, 1, $x$ (in cyclic order).

The induction step is based on a convenient application of Kuo's graphical condensation stated in Theorem {\tcb}.

\topinsert
\twoline{\mypic{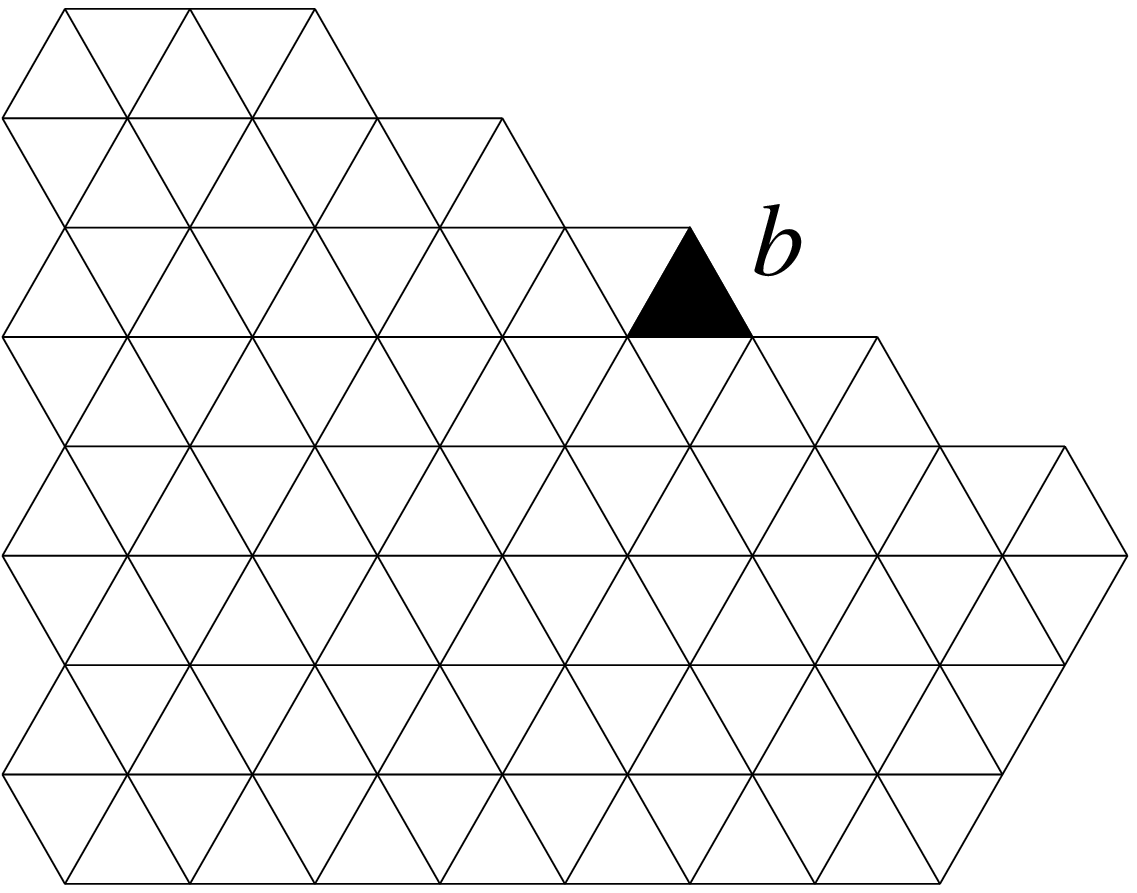}}{\mypic{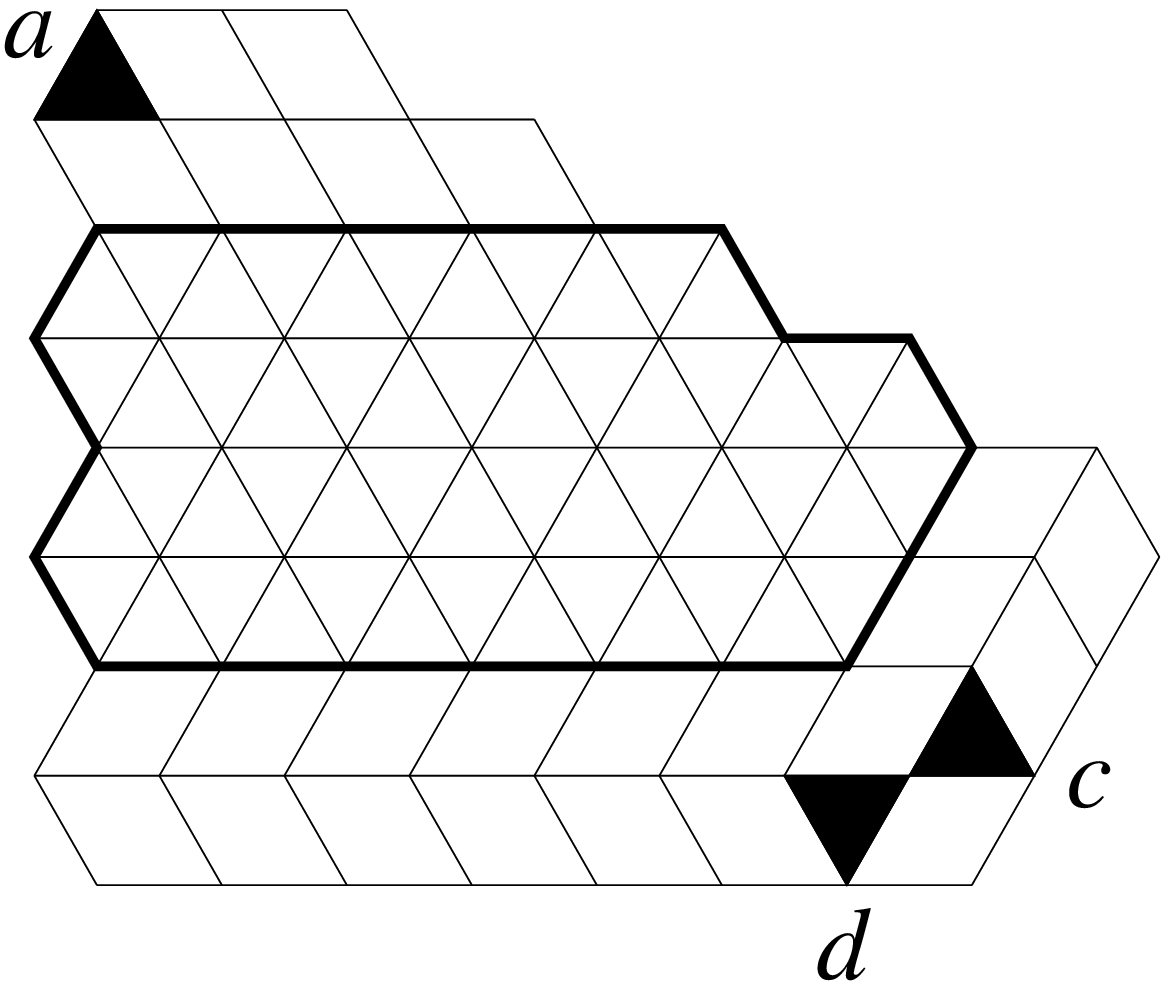}}
\bigskip

\twoline{\mypic{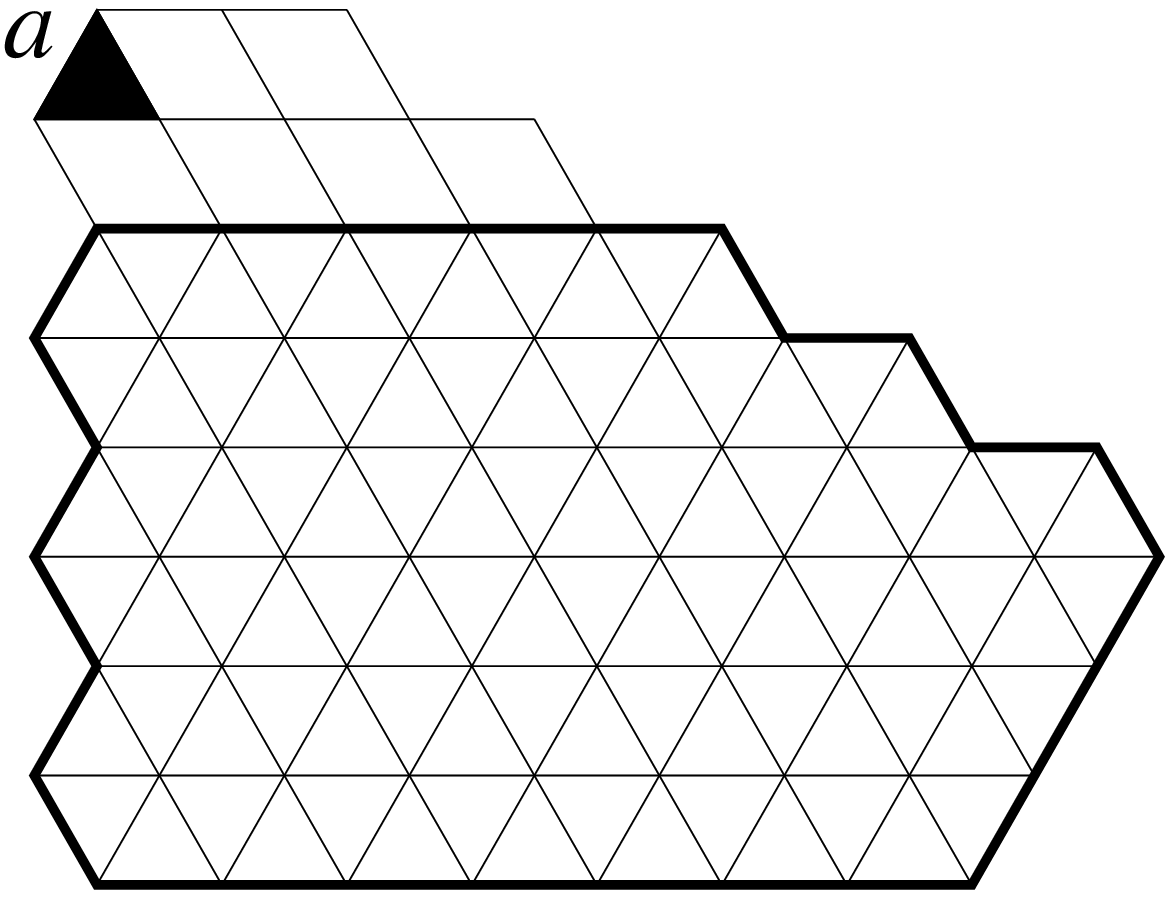}}{\mypic{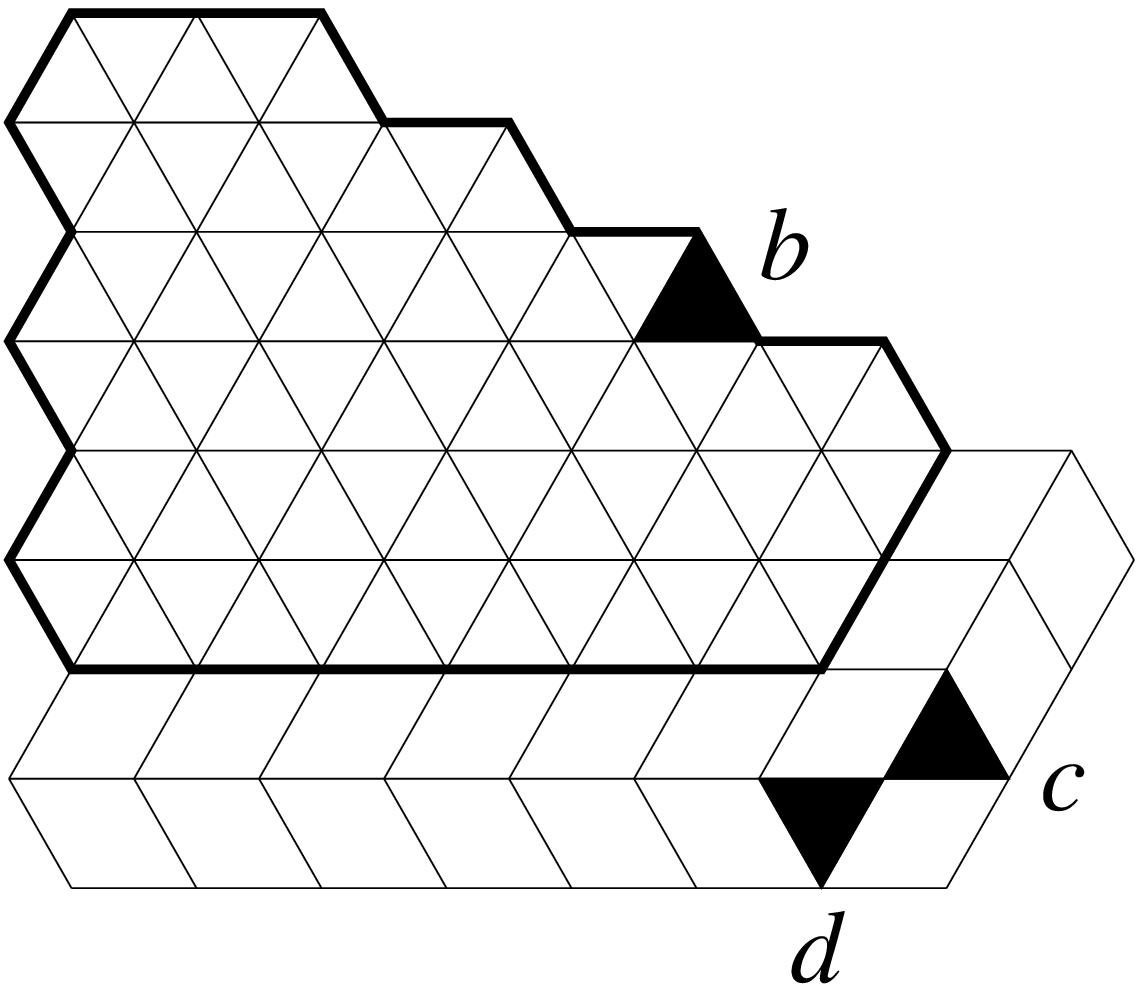}}
\bigskip

\twoline{\mypic{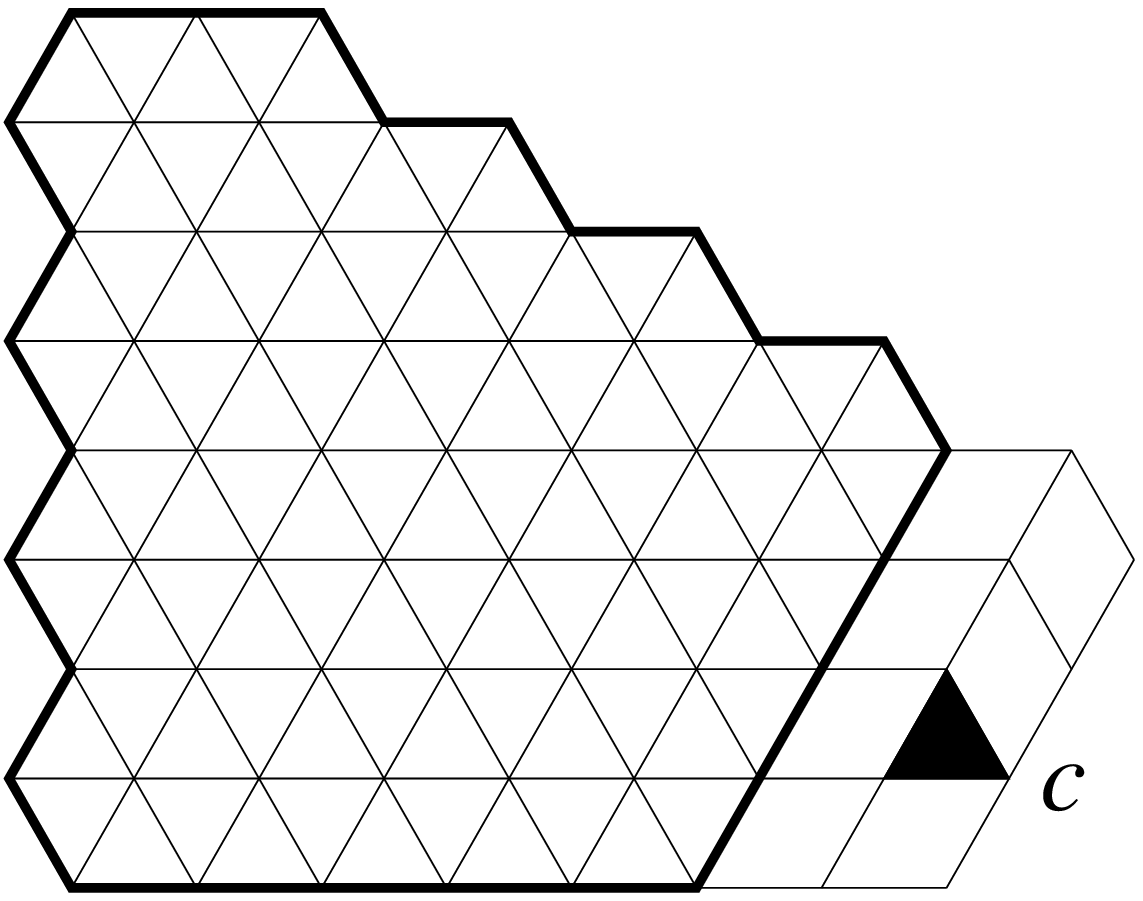}}{\mypic{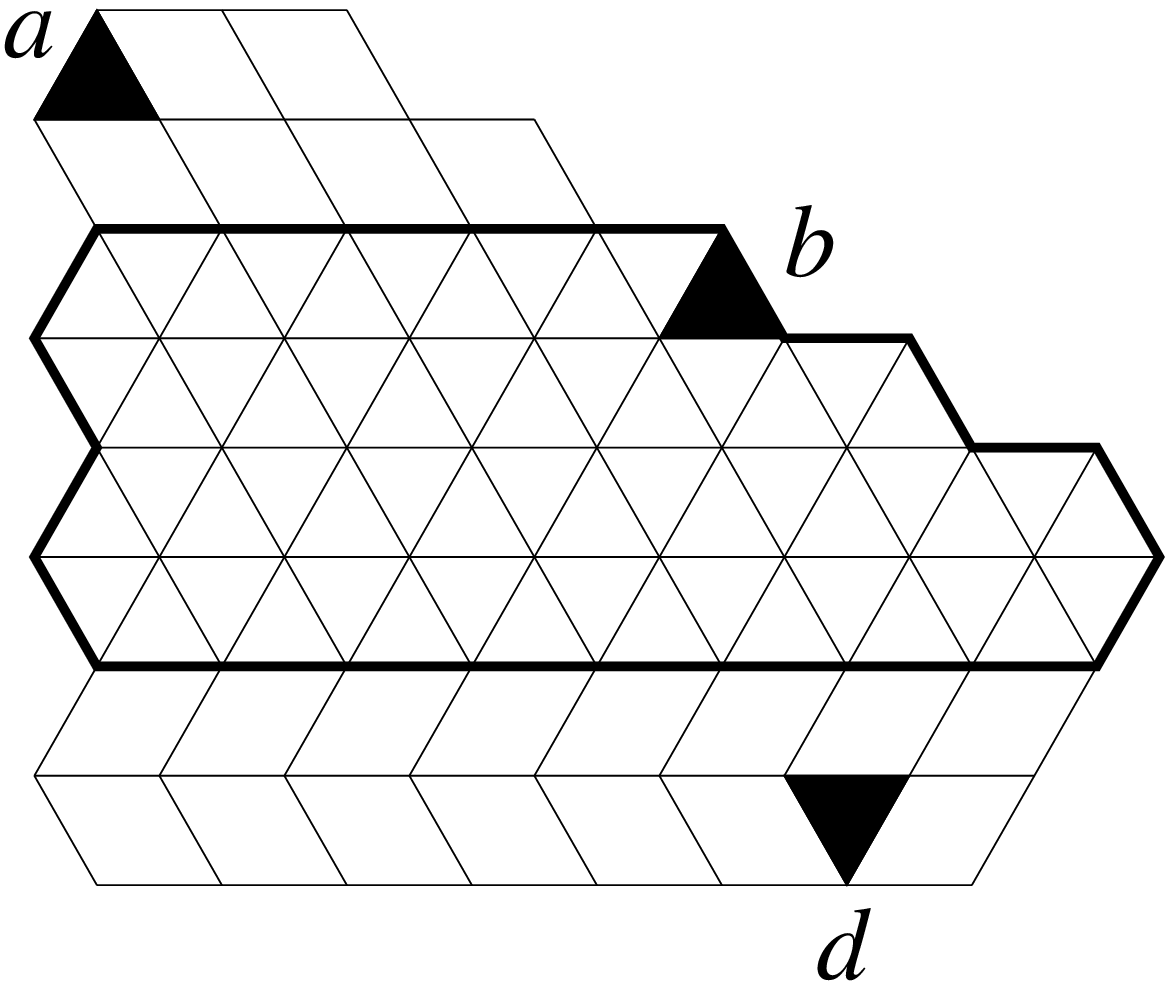}}
\bigskip
\centerline{{\smc Figure {\fcd}. {\rm The recurrence for the regions $F_{n,x}(i)$.}}}

\endinsert

\topinsert
\twoline{\mypic{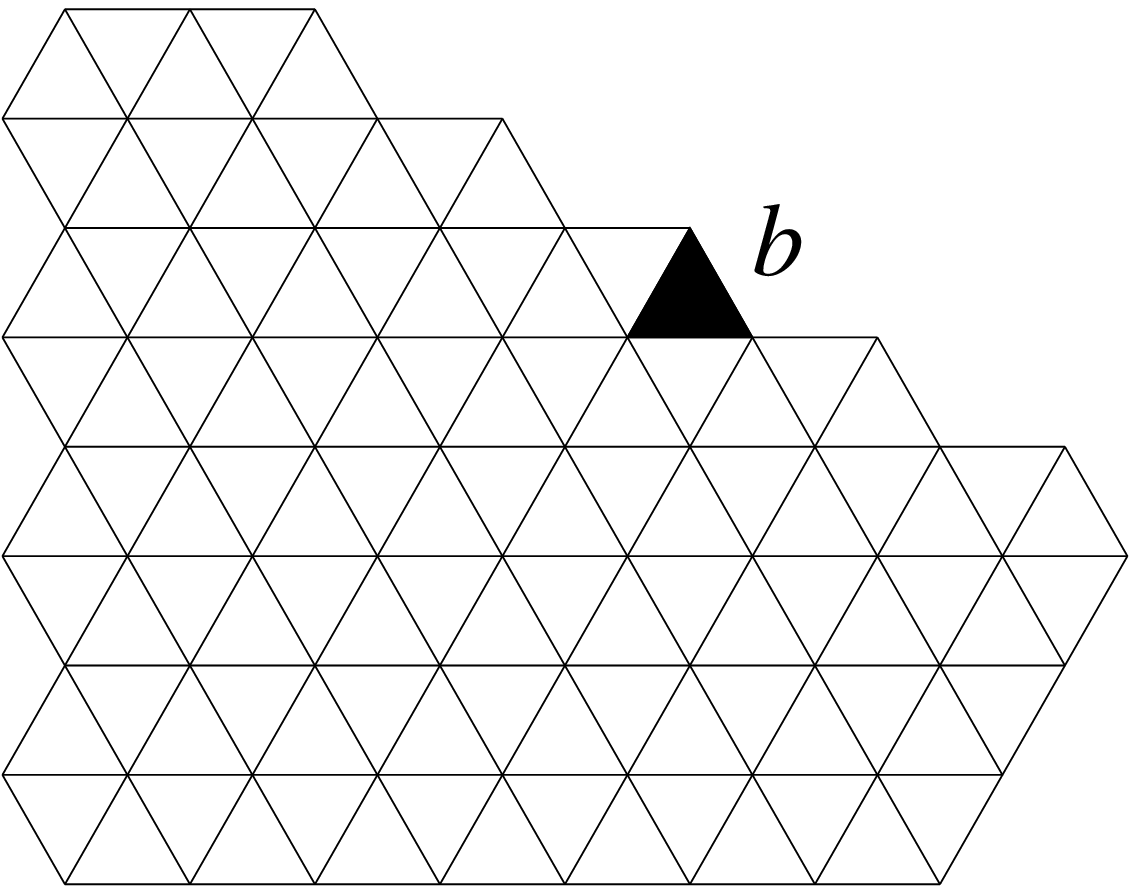}}{\mypic{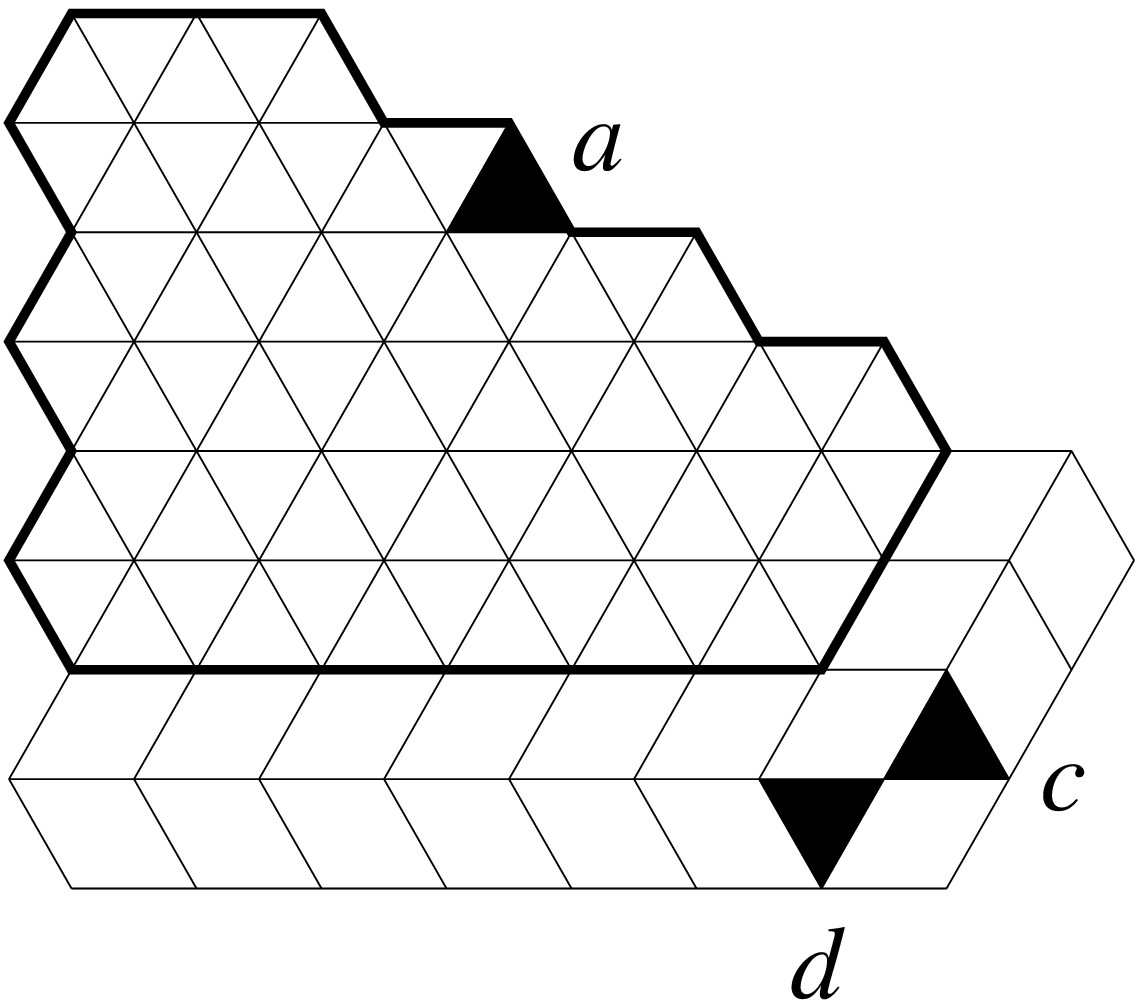}}
\bigskip

\twoline{\mypic{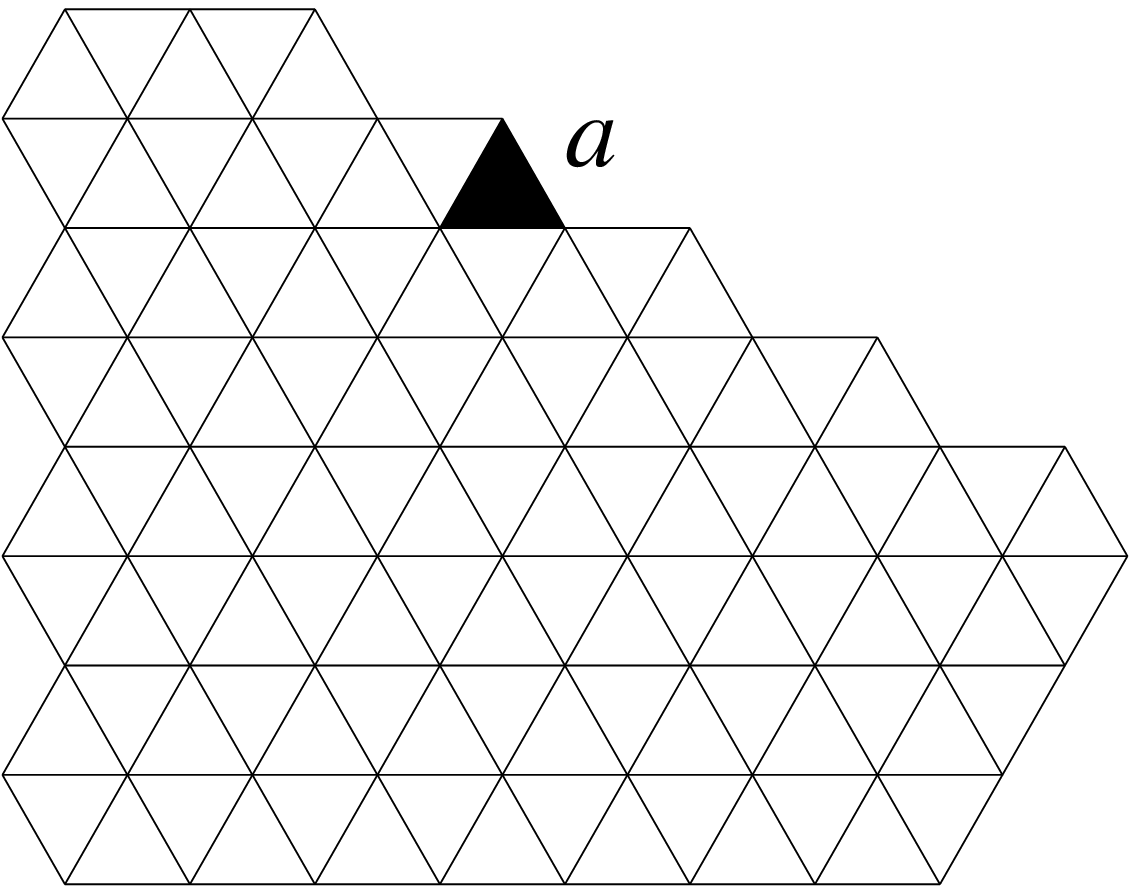}}{\mypic{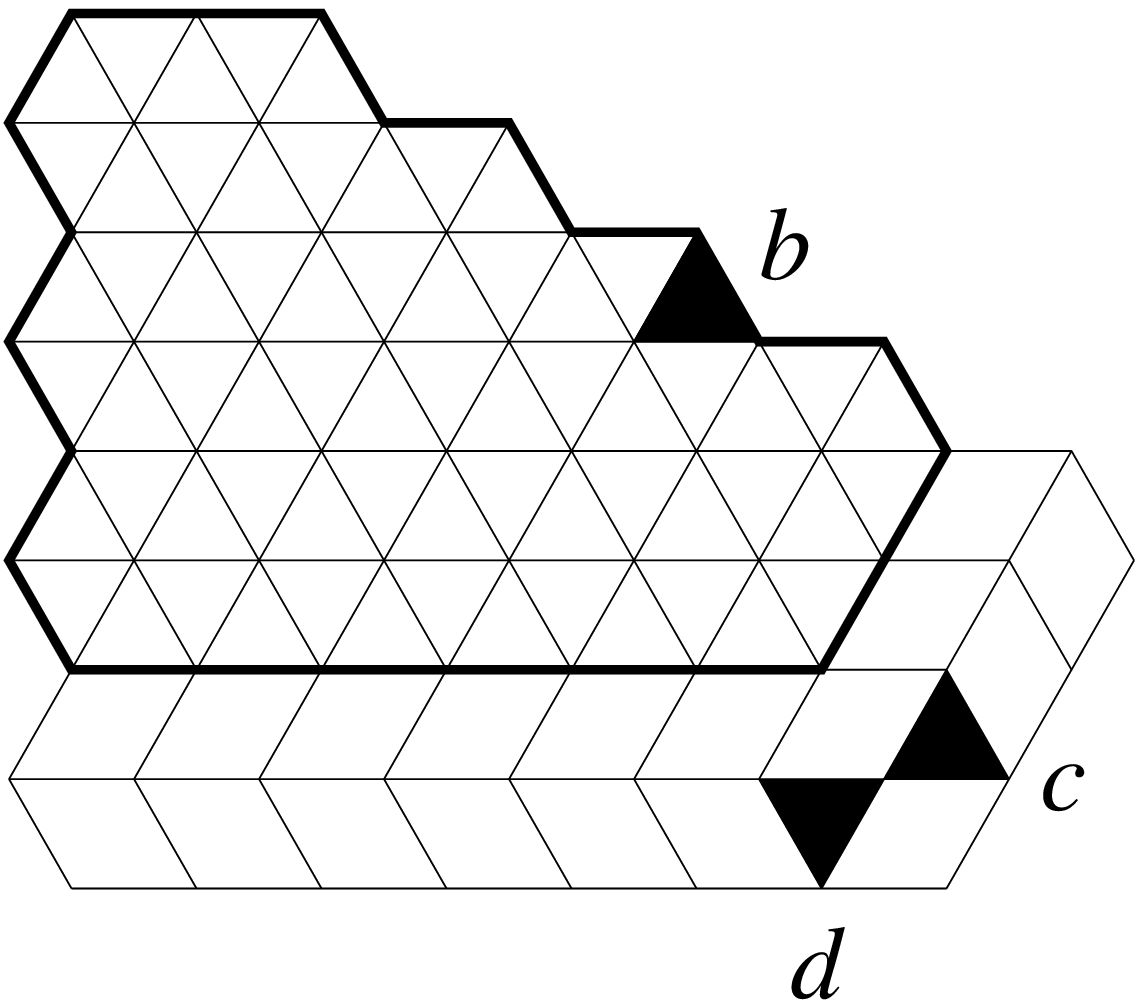}}
\bigskip

\twoline{\mypic{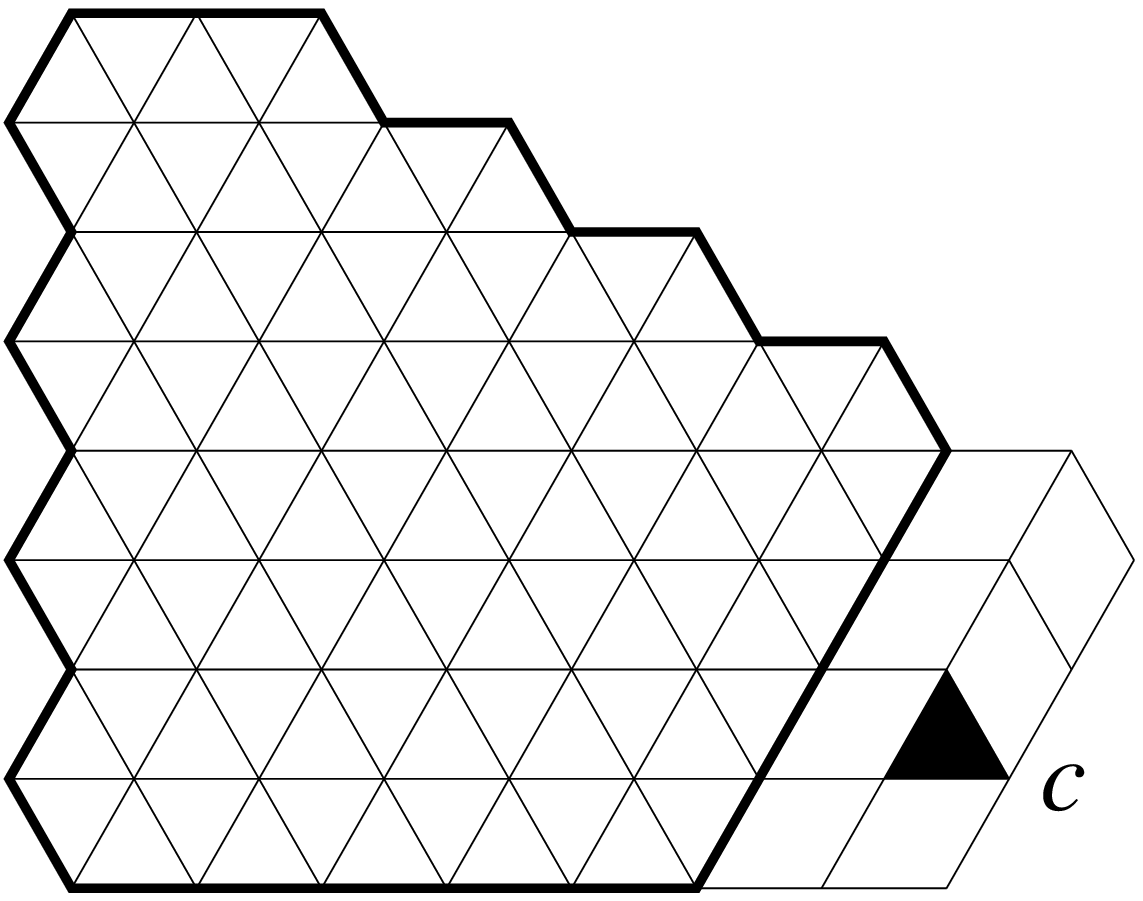}}{\mypic{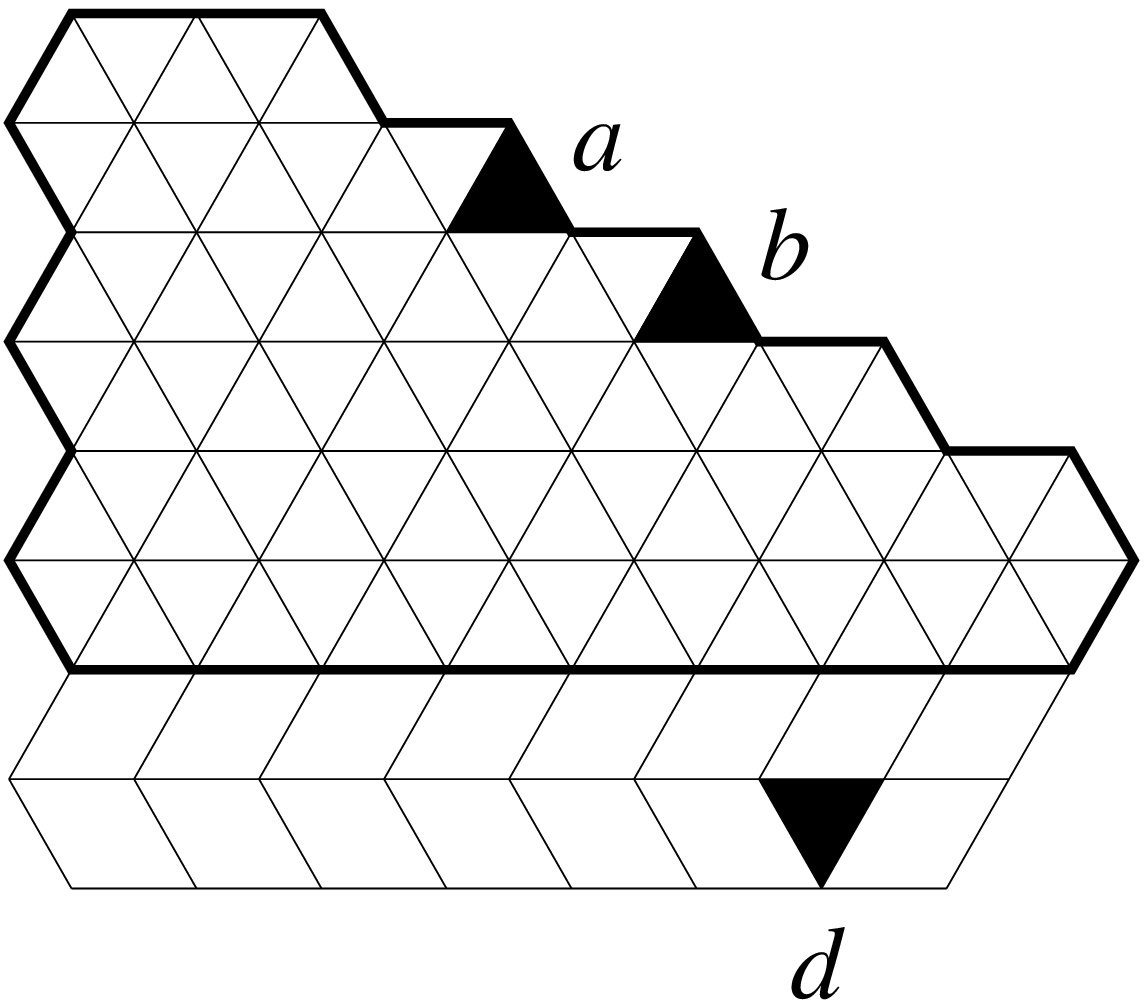}}
\bigskip
\centerline{{\smc Figure {\fce}. {\rm Reducing the $E$-regions to $F$-regions.}}}

\endinsert

Suppose $n\geq3$, and assume that (\ecd) holds for all values less than $n$. Let $G$ be the planar dual graph\footnote{By the planar dual graph of a region on the triangular lattice we understand the graph whose vertices are the unit triangles inside the region, and whose edges connect vertices corresponding to unit triangles that share an edge.} of the region obtained from $F_{n,x}(i)$ by placing back the up-pointing unit triangle that was removed from its $i$th bump. Choose the vertices $a$, $b$, $c$ and $d$ as indicated in Figure {\fcd}, where $b$ is the up-pointing unit triangle fitting in the $i$th bump from the top (Figure {\fcd} corresponds to the case $n=5$, $x=2$, $i=3$). Then (\ecf) states that the product of the number of lozenge tilings of the two regions on top is equal to the product of the number of lozenge tilings of the two regions in the middle, plus the product of the number of lozenge tilings of the two regions on the bottom. After removing the lozenges forced by the unit triangles $a$, $c$ and $d$, the leftover regions in all six instances are $F$- or $G$-type regions. More precisely, we obtain
$$
\M(F_{n,x}(i))\M(G_{n-3,x+3})=
\M(G_{n-2,x+3})\M(F_{n-1,x}(i))+\M(G_{n-1,x})\M(F_{n-2,x+3}(i-2)).\tag\ecj
$$
By the induction hypothesis, the two $F$-regions on the right hand side above have their number of lozenge tilings given by (\ecd). Using formula (\ecc) for the number of tilings of the $G$-regions, it is readily checked that the formula that results from (\ecj) for $\M(F_{n,x}(i))$ agrees with the one provided by (\ecd). This completes the induction step, and hence the proof of part (b).

(c). To prove this part we use again Kuo's graphical condensation. Choose $G$ to be the planar dual graph of the region obtained from $F_{n,x}(i)$ by placing back the up-pointing unit triangle that was removed from its $i$th bump when $F_{n,x}(i)$ was defined at the beginning of this section. Choose the vertices $a$, $b$, $c$ and $d$ as indicated in Figure {\fce}, where $a$ and $b$ are the up-pointing unit triangles in the $i$th and $j$th bumps from the top, respectively (in Figure {\fce} we have $n=5$, $x=2$, $i=2$ and $j=3$).

Then (\ecf) implies that the product of the number of lozenge tilings of the regions on the bottom is equal to the product of the number of tilings of the regions on top, minus the product of the number of tilings of the regions in the middle. After removing the lozenges forced by the unit triangles $c$ and $d$, all remaining regions turn out to be regions of type $F$ or $G$. More precisely, we obtain
$$
\M(G_{n-1,x})\M(E_{n,x}(i,j))=
\M(F_{n-1,x}(i))\M(F_{n,x}(j))-\M(F_{n-1,x}(j))\M(F_{n,x}(i)).
$$
This proves (\ece). \epf

\mysec{4. A limit formula for regions with two dents}

In our proof of Theorem {\tba} we will use the following result, which gives the limit of the ratio between the number of lozenge tilings of the regions $E_{n,1}(i,j)$ and $E_{n,1}(1,3)$, when $i$ and $j$ are fixed, and $n$ tends to infinity (the seemingly more natural choice of $E_{n,1}(1,2)$ in the denominator does not work, as $\M(E_{n,1}(1,2))=0$).

\proclaim{Proposition \tda} For any fixed integers $1\leq i<j$, we have
$$
\lim_{n\to\infty}\frac{\M(E_{n,1}(i,j))}{\M(E_{n,1}(1,3))}
=
\frac{ij(j-i)(i^2+ij+j^2-2i-2j-1)}{24}.
\tag\eda
$$

\endproclaim

\pf One readily sees that, due to forced lozenges in $F_{n,x}(1)$, one has 
$\M(F_{n,x}(1))=\M(G_{n-2,x+3})$. Therefore, (\ecd) can be rewritten as
$$
\frac{\M(F_{n,x}(i))}{\M(F_{n,x}(1))}
=
\frac{(x+1)_{i-1}\,(n-i+1)_{i-1}\,(2x+2n+i)_{i-2}(2x+2n)}
     {(2n-i)_{i-1}\,(n+x)_{i-1}\,(i-1)!}.
\tag\edb
$$
Multiplying equation (\ece) by $\M(G_{n-1,x})$ and then dividing it by $\M(F_{n-1,x}(1))\M(F_{n,x}(1))$, we obtain
$$
\spreadlines{3\jot}
\align
&
\frac{\M(E_{n,x}(i,j))\M(G_{n-1,x})}{\M(F_{n-1,x}(1))\M(F_{n,x}(1))}
=
\frac{(x+1)_{i-1}\,(n-i)_{i-1}\,(2x+2n+i-2)_{i-2}(2x+2n-2)}
     {(2n-i-2)_{i-1}\,(n+x-1)_{i-1}\,(i-1)!}
\\
&\ \ \ \ \ \ \ \ \ \ \ \ \ \ \ \ \ \ \ \ \ \ \ \ \ \ \ \ \ \ \ \ \ \ \ \ \ 
\times
\frac{(x+1)_{j-1}\,(n-j+1)_{j-1}\,(2x+2n+j)_{j-2}(2x+2n)}
     {(2n-j)_{j-1}\,(n+x)_{j-1}\,(j-1)!}
\\
&\ \ \ \ \ \ \ \ \ \ \ \ \ \ \ \ \ \ \ \ \ \ \ \ \ \ \ \ \ \ \ \ \ \ \ \ \ 
-
\frac{(x+1)_{j-1}\,(n-j)_{j-1}\,(2x+2n+j-2)_{j-2}(2x+2n-2)}
     {(2n-j-2)_{j-1}\,(n+x-1)_{j-1}\,(j-1)!}
\\
&\ \ \ \ \ \ \ \ \ \ \ \ \ \ \ \ \ \ \ \ \ \ \ \ \ \ \ \ \ \ \ \ \ \ \ \ \ 
\times
\frac{(x+1)_{i-1}\,(n-i+1)_{i-1}\,(2x+2n+i)_{i-2}(2x+2n)}
     {(2n-i)_{i-1}\,(n+x)_{i-1}\,(i-1)!}.
\tag\edc
\endalign
$$
Expressing the Pochhammer symbols as ratios of factorials, the asymptotics of each factor above follows from Stirling's formula. Setting $x=1$ in the result, and dividing by the specialization $i=1$, $j=3$, one is led to formula (\eda). \epf

\mysec{5. A double sum expression for $\omega_c(R,v)$}

The following double sum expression for the correlation $\omega_c(R,v)$ will be the starting point for our proof of Theorem {\tba}.

\proclaim{Lemma \tea} For any positive integers $R$ and $v$ we have
$$
\spreadlines{4\jot}
\align
&
\omega_c(R,v)=
\\
&
\frac{R}{24}
\left|
\sum_{a=0}^R\sum_{b=0}^R
(-1)^{a+b}
\frac{(R+a-1)!\,(R+b-1)!}{(2a)!\,(R-a)!\,(2b)!\,(R-b)!}
(b-a)^2(2v-R+a)(2v-R+b)
\right.
\\
&\ \ \ \ \ \ \ \ \ \ \ \ \ \ \ \ \ \ \ \ \ \ \ \ 
\times
\{(2v-R+a)^2+(2v-R+a)(2v-R+b)+(2v-R+b)^2
\\
&\ \ \ \ \ \ \ \ \ \ \ \ \ \ \ \ \ \ \ \ \ \ \ \ \ \ \ \ \ \ \ \ \ \ \ \ \ \ \ \,
\left.
\phantom{\sum_{a=0}^R}
-2(2v-R+a)-2(2v-R+b)-1\}\right|.
\tag\eea
\endalign
$$

\endproclaim

\topinsert
\twoline{\mypic{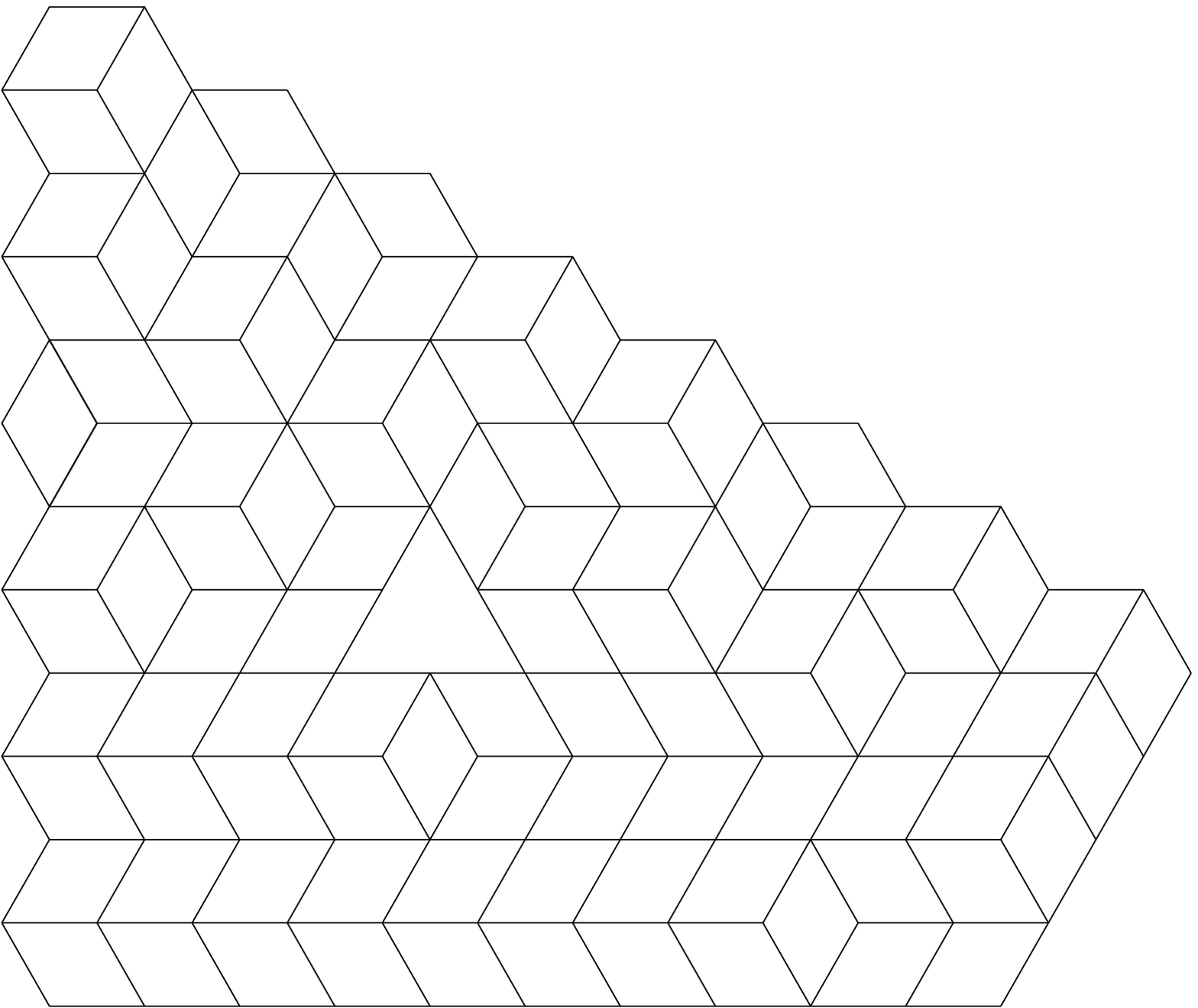}}{\mypic{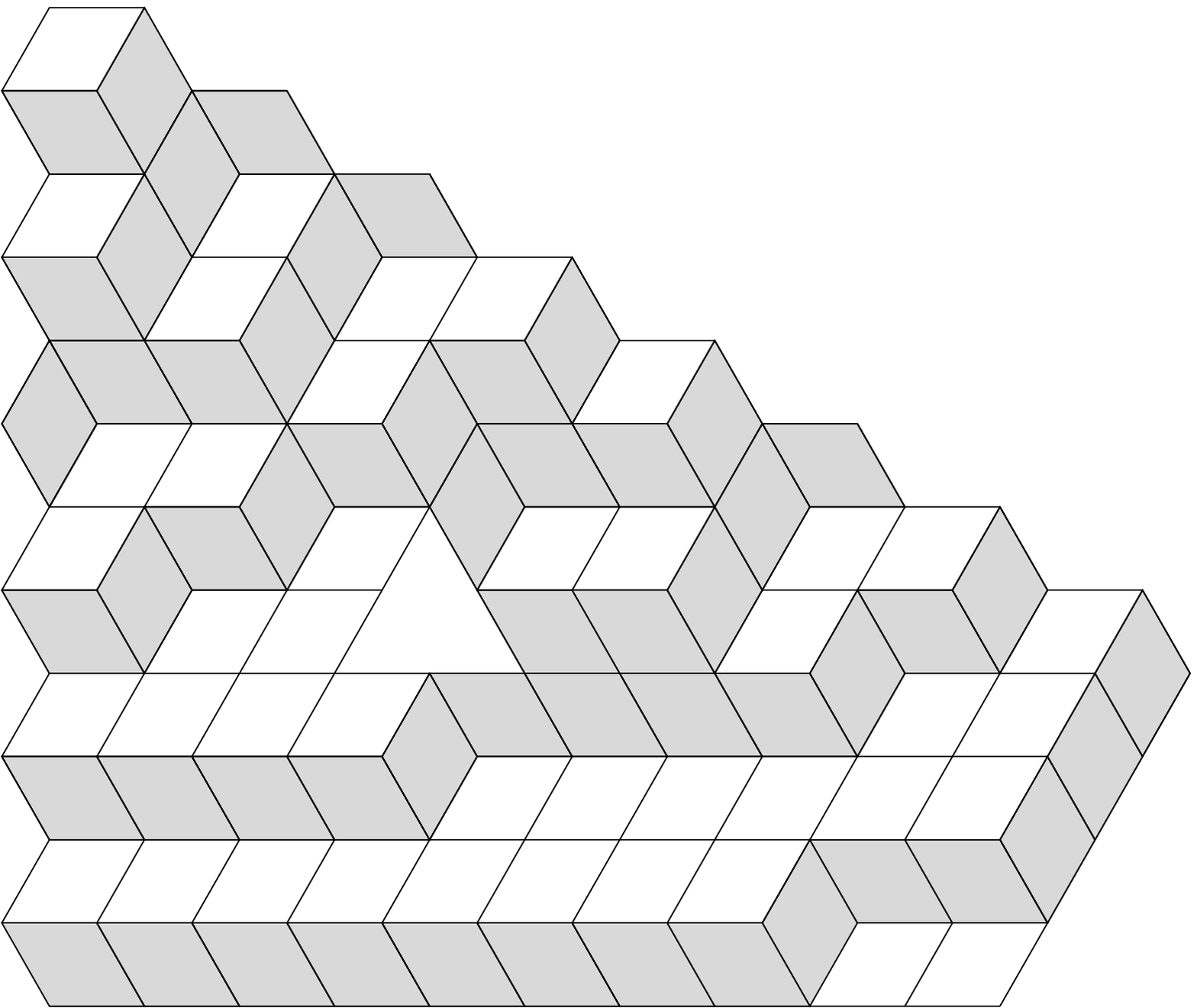}}
\medskip
\twoline{\rm A tiling of $D_{8,1}(4,4)$.}{\rm The corresponding lattice paths.}

\bigskip
\twoline{\mypic{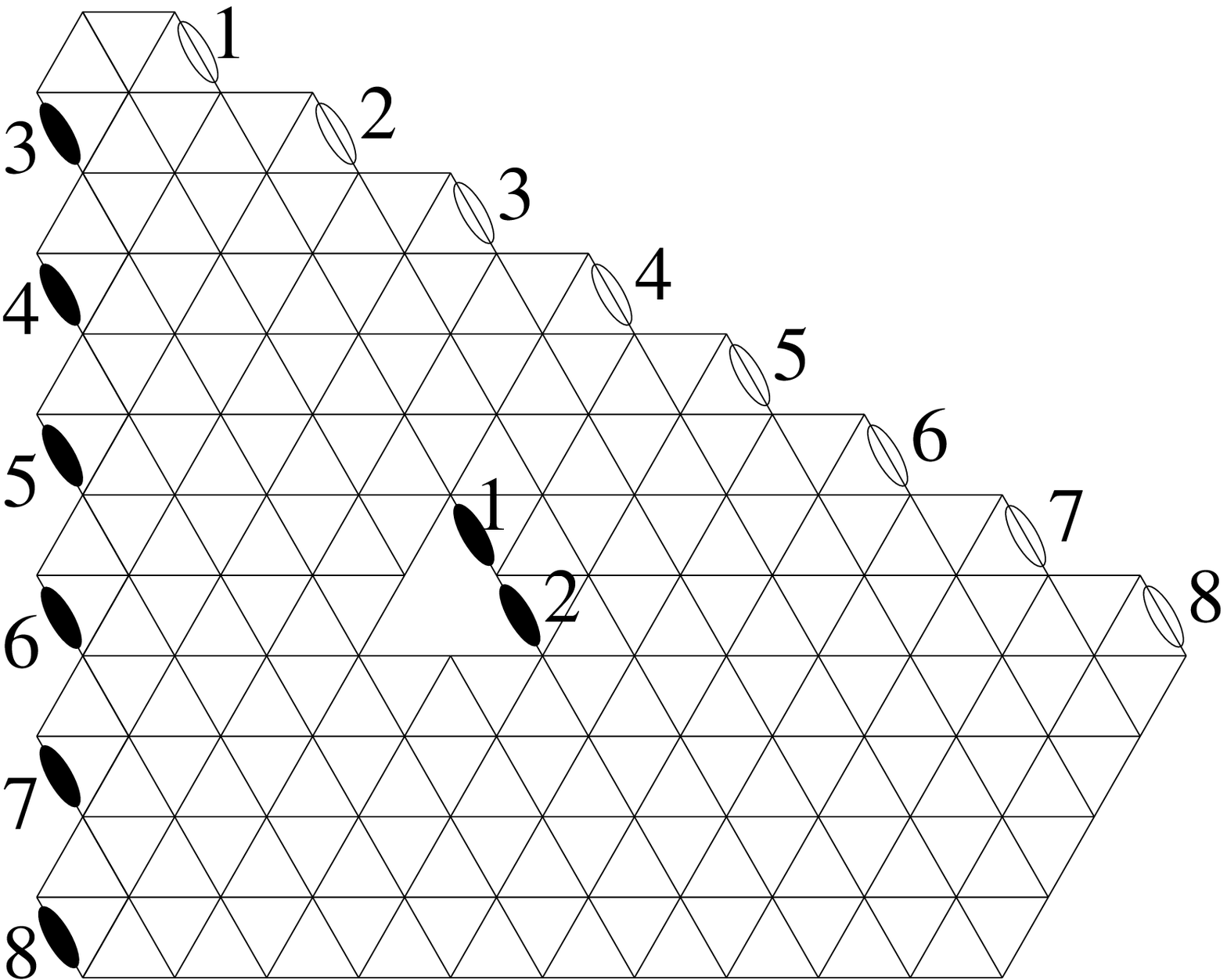}}{\mypic{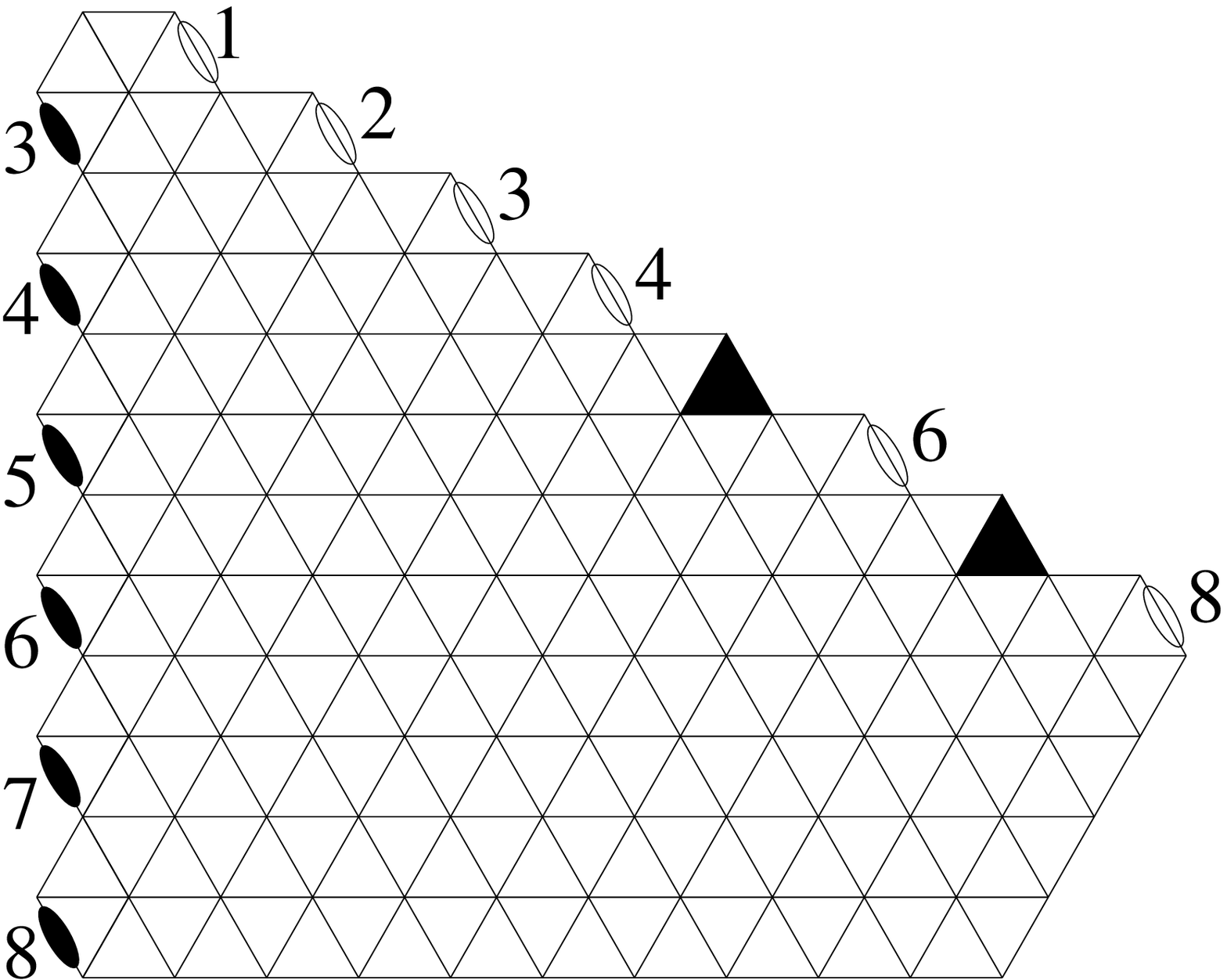}}
\medskip
\twoline{\rm Numbering of rows and columns.}{\rm $E_{8,1}(5,7)$ corresponds to a cofactor.}
\bigskip
\centerline{{\smc Figure {\fea}. {\rm Graphical Laplace expansion.}}}

\endinsert

\pf In each lozenge tiling of $D_{n,1}(R,v)$ one can see paths of unit rhombi connecting southwest facing edges on its boundary (including the gap!) to northeast facing edges on its boundary (see the top two illustrations in Figure {\fea}). According to a well-known bijection (see e.g. \cite{\DT}), these paths of rhombi uniquely determine the tiling. In turn, such families of paths of rhombi are readily seen to be in bijection with families of non-intersecting lattice paths on $\Z^2$ taking unit steps north or east, and having fixed starting and ending points. The latter, by an extension of the Gessel-Viennot-Lindstr\"om lattice path theorem (see \cite{\Lind}\cite{\GV} for the original form, and \cite{\Ste} or \cite{\sc} for the extension), can be written as a determinant. 

To be specific, let us label the starting points of the paths of rhombi starting with 1 and 2 for the two paths originating from the gap, and continuing with $3,4,\dotsc,n$ from top to bottom  for the paths starting from the western boundary (these are indicated by solid ellipses in the bottom left picture in Figure {\fea}). Label the ending points of the paths of rhombi from 1 to n, starting from the top (these are indicated by hollow ellipses in the bottom left picture in Figure {\fea}). 

Let $a_{ij}$ be the total number of paths of rhombi on the triangular lattice (with no hole in it) from starting point $i$ to ending point $j$, and set
$$
A:=((a_{ij})_{1\leq i,j\leq n}).
\tag\eec
$$
Then it follows by the above quoted results that
$$
\M(D_{n,1}(R,v))=\left|\det A\right|.
\tag\eeb
$$
Perform Laplace expansion in the above determinant along the first two rows (which recall correspond to the starting points of the two paths of rhombi originating at the gap). We obtain
$$
\det A = \sum_{1\leq a<b\leq n}(-1)^{a+b}
\det A_{\{1,2\}}^{\{a,b\}}
\det A_{[n]\setminus\{1,2\}}^{[n]\setminus\{a,b\}},
\tag\eec
$$
where $[n]$ denotes the set $\{1,\dotsc,n\}$, and for $I,J\subset[n]$, $A_I^J$ stands for the minor of $A$ corresponding to rows $I$ and columns $J$.

It is easy to see that the paths of rhombi starting at the starting points labeled 1 and 2 (i.e., starting from the gap) can only end at endpoints with labels in the range from $2v-R$ to $2v$. Therefore, for $a$ or $b$ outside this range, the matrix $A_{\{1,2\}}^{\{a,b\}}$ has a zero column, and hence its determinant is zero. Thus we can rewrite (\eec) as
$$
\det A = \sum_{0\leq a<b\leq R}(-1)^{a+b}
\det A_{\{1,2\}}^{\{2v-R+a,2v-R+b\}}
\det A_{[n]\setminus\{1,2\}}^{[n]\setminus\{2v-R+a,2v-R+b\}}.
\tag\eed
$$

It follows from the Gessel-Viennot-Lindstr\"om result (the extension is not needed here) that 
$$
\det A_{[n]\setminus\{1,2\}}^{[n]\setminus\{a,b\}}
=
\M(E_{n,1}(a,b))
\tag\eee
$$
(see the bottom right picture in Figure {\fea}).

The entries of $A$ are easily found, being given by binomial coefficients. In particular, one has
$$
\spreadlines{4\jot}
\spreadmatrixlines{4\jot}
\align
\det A_{\{1,2\}}^{\{a,b\}}
&=
\det
\left[\matrix
{R+a-1\choose 2a} & {R+b-1\choose 2b}\\
{R+a-1\choose 2a-1} & {R+b-1\choose 2b-1}
\endmatrix\right]
\\
&=
\frac{2R(b-a)(R+a-1)!\,(R+b-1)!}{(2a)!\,(R-a)!\,(2b)!\,(R-b)!}.
\tag\eef
\endalign
$$
Using (\eeb), (\eed), (\eee) and (\eef), we obtain that
$$
\spreadlines{4\jot}
\align
&
\M(D_{n,1}(R,v))=
\\
& 
\left|
2R\! \sum_{0\leq a<b\leq R}(-1)^{a+b}(b-a)
\frac{(R+a-1)!\,(R+b-1)!}{(2a)!\,(R-a)!\,(2b)!\,(R-b)!}
\M(E_{n,1}(2v-R+a,2v-R+b))\right|.
\\
\tag\eeg
\endalign
$$
Dividing the above equation by $\M(E_{n,1}(1,3))$, passing to the limit as $n\to\infty$ and using Proposition {\tda}, it follows that
$$
\spreadlines{4\jot}
\align
&
\omega_c(R,v)=
\lim_{n\to\infty}\frac{\M(D_{n,1}(R,v))}{\M(E_{n,1}(1,3))} =
\\
&
\frac{R}{12}
\left|
\sum_{0\leq a<b\leq R}(-1)^{a+b}
\frac{(R+a-1)!\,(R+b-1)!}{(2a)!\,(R-a)!\,(2b)!\,(R-b)!}
(b-a)^2(2v-R+a)(2v-R+b)
\right.
\\
&\ \ \ \ \ \ \ \ \ \ \ \ \ \ \ \ \ \ \ \ \ \ \ \ 
\times
\{(2v-R+a)^2+(2v-R+a)(2v-R+b)+(2v-R+b)^2
\\
&\ \ \ \ \ \ \ \ \ \ \ \ \ \ \ \ \ \ \ \ \ \ \ \ \ \ \ \ \ \ \ \ \ \ \ \ \ \ \ \,
\left.
\phantom{\sum_{a=0}^R}
-2(2v-R+a)-2(2v-R+b)-1\}\right|.
\tag\eeh
\endalign
$$
Note that the summand above becomes zero when $a=b$, and is invariant under swapping $a$ and $b$. This leads to (\eea). \epf

\mysec{6. Reduction of the double sum to simple sums}

Were it not for the factors
$$
\spreadlines{3\jot}
\align
&
(b-a)^2(2v-R+a)(2v-R+b)
\\
&\ \ \ \ 
\times
\{(2v-R+a)^2+(2v-R+a)(2v-R+b)+(2v-R+b)^2
\\
&\ \ \ \ \ \ \ \ \ \ \ \ \ \ \ \ \ \ \ \ \ \ \ \ \ \ \ \ \ \ \ \ \ \ \ \
-2(2v-R+a)-2(2v-R+b)-1\}
\tag\efa
\endalign
$$
in the summand, the double sum in (\eea) would separate into a product of a simple sum over $a$ and a simple sum over $b$. This is easily overcome by expanding (\efa) into a sum of monomials in $a$ and $b$:
$$
\spreadlines{3\jot}
\align
&
(b-a)^2(2v-R+a)(2v-R+b)
\\
&\ \ \ \ 
\times
\{(2v-R+a)^2+(2v-R+a)(2v-R+b)+(2v-R+b)^2
\\
&\ \ \ \ \ \ \ \ \ \ \ \ \ \ \ \ \ \ \ \ \ \ \ \ \ \ \ \ \ \ \ \ \ \ \ \
-2(2v-R+a)-2(2v-R+b)-1\}
=:
\sum_{\lambda,\mu}c_{\lambda,\mu} a^{\lambda} b^\mu,
\tag\efb
\endalign
$$
where the expression on the right hand side is a certain concrete sum of 120 monomials, which a computer algebra package can readily display; for completeness, we include it here:
$$
\spreadlines{0\jot}
\align
&
-a^3 b - 2 a^4 b + a^5 b + 2 a^2 b^2 + 2 a^3 b^2 - a^4 b^2 - a b^3 + 
 2 a^2 b^3 - 2 a b^4 - a^2 b^4 + a b^5 + a^3 R
\\
&
 + 2 a^4 R - a^5 R - 
 a^2 b R + 4 a^3 b R - 3 a^4 b R - a b^2 R - 12 a^2 b^2 R + 
 4 a^3 b^2 R + b^3 R + 4 a b^3 R 
\\
&
+ 4 a^2 b^3 R + 2 b^4 R - 3 a b^4 R -
  b^5 R - a^2 R^2 - 6 a^3 R^2 + 4 a^4 R^2 + 2 a b R^2 + 6 a^2 b R^2 + 
 2 a^3 b R^2 
\\
&
- b^2 R^2 + 6 a b^2 R^2 - 12 a^2 b^2 R^2 - 6 b^3 R^2 + 
 2 a b^3 R^2 + 4 b^4 R^2 + 4 a^2 R^3 - 6 a^3 R^3 - 8 a b R^3
\\
&
 + 
 6 a^2 b R^3 + 4 b^2 R^3 + 6 a b^2 R^3 - 6 b^3 R^3 + 3 a^2 R^4 - 
 6 a b R^4 + 3 b^2 R^4 - a^3 v - 2 a^4 v + a^5 v
\\
&
 + a^2 b v - 
 4 a^3 b v + 3 a^4 b v + a b^2 v + 12 a^2 b^2 v - 4 a^3 b^2 v - 
 b^3 v - 4 a b^3 v - 4 a^2 b^3 v - 2 b^4 v  
\\
&
+ b^5 v + 
 2 a^2 R v + 12 a^3 R v - 8 a^4 R v - 4 a b R v - 12 a^2 b R v - 
 4 a^3 b R v + 2 b^2 R v - 12 a b^2 R v 
\\
&
+ 24 a^2 b^2 R v + 
 12 b^3 R v - 4 a b^3 R v - 8 b^4 R v - 12 a^2 R^2 v + 18 a^3 R^2 v + 
 24 a b R^2 v - 18 a^2 b R^2 v
\\
&
 - 12 b^2 R^2 v - 18 a b^2 R^2 v + 
 18 b^3 R^2 v - 12 a^2 R^3 v + 24 a b R^3 v - 12 b^2 R^3 v - 
 a^2 v^2 - 6 a^3 v^2 
\\
&
+ 4 a^4 v^2 + 2 a b v^2 + 6 a^2 b v^2 + 
 2 a^3 b v^2 - b^2 v^2 + 6 a b^2 v^2 - 12 a^2 b^2 v^2 - 6 b^3 v^2 + 
 2 a b^3 v^2 + 4 b^4 v^2 
\\
&
+ 12 a^2 R v^2 - 18 a^3 R v^2 - 
 24 a b R v^2 + 18 a^2 b R v^2 + 12 b^2 R v^2 + 18 a b^2 R v^2 - 
 18 b^3 R v^2 
\\
&
+ 18 a^2 R^2 v^2 
- 36 a b R^2 v^2 + 18 b^2 R^2 v^2 - 
 4 a^2 v^3 + 6 a^3 v^3 + 8 a b v^3 - 6 a^2 b v^3 - 4 b^2 v^3 - 
 6 a b^2 v^3 
\\
&
+ 6 b^3 v^3 
- 12 a^2 R v^3 + 24 a b R v^3 - 
 12 b^2 R v^3 + 3 a^2 v^4 - 6 a b v^4 + 3 b^2 v^4+ 3 a b^4 v.
\tag\efc
\endalign
$$
Then we obtain from Lemma {\tea} that
$$
\spreadlines{4\jot}
\align
\omega_c(R,v)
&=
\frac{1}{24 R}
\left|
\sum_{a=0}^R\sum_{b=0}^R
(-1)^{a+b}
\frac{(R+a-1)!\,(R+b-1)!}{(2a)!\,(R-a)!\,(2b)!\,(R-b)!}
\left\{\sum_{\lambda,\mu}c_{\lambda,\mu} a^{\lambda} b^\mu\right\}\right|
\\
&=
\frac{1}{24 R}
\left|
\sum_{\lambda,\mu}c_{\lambda,\mu}
\left(\sum_{a=0}^R (-1)^a\frac{(R+a-1)!}{(2a)!\,(R-a)!}a^\lambda\right)
\left(\sum_{b=0}^R (-1)^b\frac{(R+b-1)!}{(2b)!\,(R-b)!}b^\mu\right)
\right|.
\\
\tag\efd
\endalign
$$
We can conveniently express the resulting sums in terms of hypergeometric functions\footnote{The hypergeometric function of parameters $a_1,\dotsc,a_p$ and $b_1,\dotsc,b_q$ is defined by
$$
{}_p F_q\left[\matrix a_1,\dotsc,a_p\\b_1,\dotsc,b_q\endmatrix;z\right]:=
\sum_{i=0}^\infty\frac{(a_1)_i\cdots(a_p)_i}{i!\,(b_1)_i\cdots(b_q)_i}z^i.
$$}
and their moments. Indeed, writing
$$
\frac{(R+a-1)!}{(2a)!\,(R-a)!}
=
\frac{(-1)^a}{R}\frac{(-R)_a\,(R)_a}{4^a (1)_a\,(1/2)_a},
$$
we obtain
$$
\spreadlines{3\jot}
\align
\sum_{a=0}^R (-1)^a\frac{(R+a-1)!}{(2a)!\,(R-a)!}
&=
\frac{1}{R}
\sum_{a=0}^R 
\frac{(-R)_a\,(R)_a}{(1)_a\,(1/2)_a}\left(\frac{1}{4}\right)^a
\\
&={}_2 F_1\left[\matrix -R,\,R\\ \ \ 1/2\endmatrix;\frac14\right].
\tag\efe
\endalign
$$
For $k$ a non-negative integer, define the $k$th moment of the above hypergeometric function by
$$
S^{(k)}(R;x):=
\sum_{i=0}^R \frac{(-R)_i\,(R)_i}{(1)_i\,(1/2)_i} x^i i^k.
\tag\eff
$$
Then from (\efd) we obtain the following result.

\proclaim{Proposition \tfa} We have
$$
\omega_c(R,v)
=
\frac{1}{24 R}
\left|\sum_{\lambda,\mu}c_{\lambda,\mu}S^{(\lambda)}(R;1/4)
S^{(\mu)}(R;1/4)\right|,
\tag\efg
$$
where the $c_{\lambda,\mu}$'s are given by ${\text{\rm (\efb)}}$, and the moments $S^{(k)}(R;x)$ by ${\text{\rm (\eff)}}$. \epf

\endproclaim

\mysec{7. Computation of the moments $S^{(k)}(R;1/4)$}

As the largest exponent of $a$ or $b$ in (\efc) is 5, in order to evaluate the sum (\efg) we need to find the moments $S^{(k)}(R;1/4)$ for $k=0,1,\dotsc,5$. These are provided by the following result.

\proclaim{Lemma \tga} For any non-negative integer $R$ we have
$$
\spreadlines{3\jot}
\align
S^{(0)}(R;1/4)&=\cos\left(\frac{R\pi}{3}\right)
\tag\ega\\
S^{(1)}(R;1/4)&=-\frac{R}{\sqrt{3}}\sin\left(\frac{R\pi}{3}\right)
\tag\egb\\
S^{(2)}(R;1/4)&=-\frac{R^2}{3}\cos\left(\frac{R\pi}{3}\right)
-\frac{2R}{3\sqrt{3}}\sin\left(\frac{R\pi}{3}\right)
\tag\egc\\
S^{(3)}(R;1/4)&=\frac{R(R^2-2)}{3\sqrt{3}}\sin\left(\frac{R\pi}{3}\right)
-\frac{2R^2}{3}\cos\left(\frac{R\pi}{3}\right)
\tag\egd\\
S^{(4)}(R;1/4)&=\frac{R^2(R^2-12)}{9}\cos\left(\frac{R\pi}{3}\right)
+\frac{2R(6R^2-5)}{9\sqrt{3}}\sin\left(\frac{R\pi}{3}\right)
\tag\ege\\
S^{(5)}(R;1/4)&=\frac{-R(3R^4-120R^2+74)}{27\sqrt{3}}\sin\left(\frac{R\pi}{3}\right)
+\frac{10R^2(2R^2-9)}{27}\cos\left(\frac{R\pi}{3}\right)
\tag\egf
\endalign
$$

\endproclaim

\pf Define the $k$th {\it descending} moments $S^{\langle k\rangle}(R;x)$ by
$$
S^{\langle k\rangle}(R;x):=
\sum_{i=0}^R \frac{(-R)_i\,(R)_i}{(1)_i\,(1/2)_i} x^i i(i-1)\cdots(i-k+1).
\tag\egg
$$
It readily follows by differentiating $S^{\langle 0\rangle}(R;x)=
{}_2 F_1\left[\matrix -R,\,R\\ \ 1/2\endmatrix;x\right]$ $k$ times that
$$
S^{\langle k\rangle}(R;x)=
x^k\frac{d^k}{dx^k}\,{}_2 F_1\left[\matrix -R,\,R\\ \ 1/2\endmatrix;x\right].
\tag\egh
$$
On the other hand, by a well known formula (see e.g. \cite{\PBM}), we have the closed form evaluation
$$
{}_2 F_1\left[\matrix -R,\,R\\ \ 1/2\endmatrix;x\right]
=
\cos(2R\arcsin\sqrt{x}),
\tag\egi
$$
which upon specializing $x=1/4$ gives (\ega).

Using (\egi), we obtain from (\egh), after simplifications, that
$$
\spreadlines{3\jot}
\align
S^{\langle 1\rangle}(R;1/4)&=-\frac{R}{\sqrt{3}}\sin\left(\frac{R\pi}{3}\right)
\tag\egj\\
S^{\langle 2\rangle}(R;1/4)&=-\frac{R^2}{3}\cos\left(\frac{R\pi}{3}\right)
+\frac{R}{3\sqrt{3}}\sin\left(\frac{R\pi}{3}\right)
\tag\egk\\
S^{\langle 3\rangle}(R;1/4)&=\frac{R(R^2-2)}{3\sqrt{3}}\sin\left(\frac{R\pi}{3}\right)
+\frac{R^2}{3}\cos\left(\frac{R\pi}{3}\right)
\tag\egl\\
S^{\langle 4\rangle}(R;1/4)&=\frac{R^2(R^2-9)}{9}\cos\left(\frac{R\pi}{3}\right)
-\frac{2R(3R^2-7)}{9\sqrt{3}}\sin\left(\frac{R\pi}{3}\right)
\tag\egm\\
S^{\langle 5\rangle}(R;1/4)&=\frac{-R(3R^4-75R^2+152)}{27\sqrt{3}}\sin\left(\frac{R\pi}{3}\right)
-\frac{10R^2(R^2-9)}{27}\cos\left(\frac{R\pi}{3}\right)
\tag\egn
\endalign
$$
Clearly, the moments $S^{(1)}$ and $S^{\langle 1\rangle}$ are the same. The higher order moments (\eff) can be easily expressed as linear combinations of the descending moments (\egg)
using the relations
$$
\spreadlines{3\jot}
\align
i^2&=i(i-1)+i\\
i^3&=i(i-1)(i-2)+3i(i-1)+i\\
i^4&=i(i-1)(i-2)(i-3)+6i(i-1)(i-2)+7i(i-1)+i\\
i^5&=i(i\!-\!1)(i\!-\!2)(i\!-\!3)(i\!-\!4)+10i(i\!-\!1)(i\!-\!2)(i\!-\!3)+25i(i\!-\!1)(i\!-\!2)+15i(i\!-\!1)+i
\endalign
$$
The corresponding linear combinations of equalities (\egj)--(\egm) yield (\egb)--(\egf). \epf

\mysec{8. Proof of Theorem {\tba}}

Here we put together the results from the previous sections to prove our main result.

{\it Proof of Theorem {\tba}.} By Proposition {\tfa}, it suffices to evaluate the sum $\Cal S$ on the right hand side of (\efg), where $c_{\lambda,\mu}$ is the coefficient of $a^\lambda b^\mu$ in (\efc). Thus $\Cal S$ is a sum of 120 terms, each equal to a concrete monomial in $R$ and $v$ times the product of two moments from the list (\ega)--(\egf). Using a computer algebra package such as Mathematica, this yields, by Proposition {\tfa}, that
$$
\spreadlines{3\jot}
\align
&
{\Cal S}
=
-\frac{2}{27}R^2(2R-6v-1)(4R-6v-1)[4R^2-2R(6v+1)+4v(3v+1)-2]
\\
&
+\frac{2}{27}\cos\left(\frac{2R \pi}{3}\right)
R(2R-4v-1)[6R^2-3R(12v+1)+36v^2+6v+1]
\\
&
-\frac{\sqrt{3}}{81}\sin\left(\frac{2R \pi}{3}\right)
[24R^4-24R^3-2R^2(216v^2+12v-5)-2R(12v+1)(36v^2+6v-1)
\\
&\ \ \ \ \ \ \ \ \ \ \ \ \ \ \ \ \ \ \ \ \ \ \ \ \ \ \ \ \ \ \ \ \ \ \ \ \ \ \ \ \ \ \ \ \ \ \ \ \ 
\phantom{\frac{2}{27}}
-4v(3v+1)(6v-1)(6v+1)].
\tag\eha
\endalign
$$

For $R$ a multiple of 3, this simplifies to
$$
{\Cal S}
=
-\frac{8}{27}
R^2(R-3v)(2R-3v)(4R^2-12Rv+12v^2-8R+16v+3).
\tag\ehb
$$
When $R$ is a multiple of 3 plus 1, (\eha) becomes
$$
{\Cal S}
=
-\frac{4}{27}
R(2R+1)(R-3v-1)(4R-6v-1)(2R^2-6Rv+6v^2-R-v),
\tag\ehc
$$
while for multiples of 3 minus 1 it is
$$
{\Cal S}
=
-\frac{4}{27}
R(2R-1)(2R-6v-1)(2R-3v-1)(2R^2-6Rv+6v^2+2R-v).
\tag\ehd
$$
By Proposition {\tfa}, this implies that the exact value of the correlation $\omega_c(R,v)$ is
$$
\spreadlines{3\jot}
\align
&
\omega_c(R,v)=
\\
&
\left\{\matrix 
\frac{1}{81}R(R-3v)(2R-3v)(4R^2-12Rv+12v^2-8R+16v+3),\  R=0\, (\text{\rm mod}\, 3)
\\
\\
\frac{1}{162}
(2R+1)(R-3v-1)(4R-6v-1)(2R^2-6Rv+6v^2-R-v),\  R=1\, (\text{\rm mod}\, 3)
\\
\\
\!\!
\frac{1}{162}
(2R-1)(2R-6v-1)(2R-3v-1)(2R^2-6Rv+6v^2+2R-v),\  R=2\, (\text{\rm mod}\, 3)
\endmatrix
\right.,
\\
\tag\ehe
\endalign
$$
which proves the first asymptotic equality in (\ebb).

In order to prove the second asymptotic equality, consider a rectangular coordinate system in Figure {\fbe} with origin at the intersection of the two dotted lines $\ell_1$ and $\ell_2$. It is readily verified that the coordinates of the midpoint of the basis of the original gap $O_1$ are $(3v-2R,-v\sqrt{3})$, while the coordinates of its mirror images in $O_2,\dotsc,O_5$ are 
$$
\spreadlines{3\jot}
\align
&
(3v-R,R\sqrt{3}-v\sqrt{3}),\
(2R-3v, -v\sqrt{3}),\
(R-3v,R\sqrt{3}-v\sqrt{3}),\
(R,2v\sqrt{3}-R\sqrt{3}),
\\
&\text{\rm and}\ (-R,2v\sqrt{3}-R\sqrt{3}),
\endalign
$$
respectively. Therefore we have
$$
\spreadlines{3\jot}
\align
\de(O_1,O_2)&=2R
\\
\de(O_1,O_3)&=2(3v-2R)
\\
\de(O_1,O_4)&=\de(O_1,O_5)=2\sqrt{3(R^2-3Rv+3v^2)}
\\
\de(O_1,O_6)&=2(3v-R),
\endalign
$$
which implies the second asymptotic equality in (\ebb). \epf

\mysec{9. Concluding remarks}

We have analyzed the interaction of a gap with a $60^\circ$ corner in a sea of dimers on the hexagonal lattice, and obtained the exact expressions (\ehe) for the resulting correlation. The asymptotics of this correlation turns out to be given, up to a multiplicative constant, by the sixth root of the joint correlation of the collection of six gaps consisting of the original gap and its five mirror images in the sides of the $60^\circ$ angle. This constitutes a more complex situation in which electrostatic analogs apply in the context of correlations of gaps in dimer systems (two related situations are treated in \cite{\sc} and \cite{\free}, but they concern the interaction with an edge --- as opposed to a corner --- and give rise to a single image for a gap). 

We conclude this paper by pointing out the interesting feature that even before taking asymptotics, the exact values of the correlation $\omega_c(R,v)$ already present a high degree of factorization. It would be interesting to understand why this is the case, as well as any physical interpretation that this might have.

\mysec{References}
{\openup 1\jot \frenchspacing\raggedbottom
\roster

\myref{\And}
  G. E. Andrews, Plane partitions (III): The weak Macdonald
conjecture, {\it Invent. Math.} {\bf 53} (1979), 193--225.

\myref{\fiveandahalf}
M. Ciucu and C. Krattenthaler, Plane partitions II: $5\frac {1}{2}$
symmetry classes,
in: ``Combinatorial Methods in Representation
Theory,'' M.~Kashiwara, K.~Koike, S.~Okada, I.~Terada, H.~Yamada, eds., RIMS, Kyoto,
{\it Advanced Study in Pure Mathematics} {\bf 28} (2000), 83-103.  

\myref{\ri}
  M. Ciucu, Rotational invariance of quadromer correlations on the hexagonal lattice, \
{\it Adv. in Math.} {\bf 191} (2005), 46-77.

\myref{\sc}
  M. Ciucu, A random tiling model for two dimensional electrostatics, {\it Mem. Amer. \
Math. Soc.} {\bf 178} (2005), no. 839, 1--106.

\myref{\ec}
  M. Ciucu, The scaling limit of the correlation of holes on the triangular lattice
with periodic boundary conditions, {\it Mem. Amer. Math. Soc.} {\bf 199} (2009),
no. 935, 1-100.

\myref{\ef}
  M. Ciucu, The emergence of the electrostatic field as a Feynman sum in random
tilings with holes, {\it Trans. Amer. Math. Soc.} {\bf 362} (2010), 4921-4954.

\myref{\ov}
  M. Ciucu, Dimer packings with gaps and electrostatics, {\it Proc. Natl. Acad. Sci.
USA} {\bf 105} (2008), 2766-2772.

\myref{\free}
  M. Ciucu and C. Krattenthaler, The interaction of a gap with a free boundary in a
two dimensional dimer system, {\it Comm. Math. Phys.} {\bf 302} (2011), 253-289.

\myref{\gd}
  M. Ciucu, The interaction of collinear gaps of arbitrary charge in a two
dimensional dimer system, 2012, arXiv:1202.1188.

\myref{\DT}
  G. David and C. Tomei, The problem of the calissons,
{\it Amer\. Math\. Monthly} {\bf 96} (1989), 429--431.

\myref{\FS}
  M. E. Fisher and J. Stephenson, Statistical mechanics of dimers on a plane
lattice. II. Dimer correlations and monomers, {\it Phys. Rev. (2)} {\bf 132} (1963),
1411--1431.

\myref{\GV}
  I. M. Gessel and X. Viennot, Binomial determinants, paths, and hook length formulae,
{\it Adv. in Math.} {\bf 58} (1985), 300--321.

\myref{\Kuo}
Kuo, Eric H. Applications of graphical condensation for enumerating matchings and tilings. Theoret. Comput. Sci. 319 (2004), no. 1-3, 29–57.

\myref{\Kup}
  G. Kuperberg, Symmetries of plane partitions and the permanent-de\-ter\-mi\-nant
method, {\it J. Combin. Theory Ser. A} {\bf 68} (1994), 115--151.


\myref{\Lind}
  B. Lindstr\"om, On the vector representations of induced
matroids, {\it Bull. London Math. Soc.} {\bf 5} (1973), 85--90.

\myref{\MRR}
  W. H. Mills, D. P. Robbins and H. Rumsey Jr., Enumeration of a symmetry class of
plane partitions, {\it Discrete Math.} {\bf 67} (1987), 43--55.

\myref{\PBM}
  A.P. Prudnikov, Yu.A. Brychkov and O.I. Marichev, ``Integrals and series,'' vol. 3,
Gordon and Breach Science Publishers, New York, 1986.

\myref{\Sta}
  R. P. Stanley, Ordered structures and partitions, {\it Memoirs of the Amer. Math. Soc.,} 
no. 119 (1972).

\myref{\Ste}
  J. R. Stembridge, Nonintersecting paths, Pfaffians and plane partitions,
{\it Adv. in Math.} {\bf 83} (1990), 96--131.

\endroster\par}

\enddocument